\documentclass[12pt]{iopart}
\pdfoutput=1
\usepackage[pdftex]{graphics}
\usepackage{color}

\usepackage{iopams}
\usepackage{graphicx} 
\usepackage{relsize}
\def\beqra{\begin{eqnarray}}
\def\eeqra{\end{eqnarray}}
\def\beq{\begin{equation}}
\def\eeq{\end{equation}}
\def\Mpl{M_{\rm pl}}
\def\Hds{H_{\rm dS}}
\def\dphids{\dot{\phi}_{\rm dS}}
\def\xds{x_{\rm dS}}
\def\dphi{\dot{\phi}}
\def\ddphi{\ddot{\phi}}
\def\a{\alpha}
\def\b{\beta}
\def\g{\gamma}
\def\e{\eta}
\def\lcdm{\Lambda\textrm{CDM}}

\begin{document}

\title{Spherical Collapse in covariant Galileon theory}
\author{Emilio Bellini}
\address{Dipartimento di Fisica e Astronomia ``G. Galilei'', Universit\`a degli Studi di Padova, via Marzolo 8, I-35131, Padova, Italy\\
INFN, Sezione di Padova, via Marzolo 8, I-35131, Padova, Italy}
\ead{emilio.bellini@pd.infn.it}
\author{Nicola Bartolo}
\address{Dipartimento di Fisica e Astronomia ``G. Galilei'', Universit\`a degli Studi di Padova, via Marzolo 8, I-35131, Padova, Italy\\
INFN, Sezione di Padova, via Marzolo 8, I-35131, Padova, Italy}
\ead{nicola.bartolo@pd.infn.it}
\author{Sabino Matarrese}
\address{Dipartimento di Fisica e Astronomia ``G. Galilei'', Universit\`a degli Studi di Padova, via Marzolo 8, I-35131, Padova, Italy\\
INFN, Sezione di Padova, via Marzolo 8, I-35131, Padova, Italy}
\ead{sabino.matarrese@pd.infn.it}

\begin{abstract}

In this paper we study the evolution of a spherical matter overdensity in the context of the recently introduced Galileon field theory. 
Our analysis considers the complete covariant Lagrangian in four dimensions. This theory is composed by a potential and a standard kinetic term, 
a cubic kinetic term and two 
additional terms that include the coupling between the Galileon and the metric, to preserve the original properties of Galileons also in curved space-times. 
Here we extend previous studies, which considered both the quintessence and the cubic terms, by focussing on the role of the last two terms. 
The background evolution we consider is driven by a tracker solution. Studying scalar perturbations in the non-linear regime, 
we find constraints on the parameter of the model. We will show how the new terms contribute to the collapse phase 
and how they modify physical parameters, such as the linearized density contrast and the virial overdensity. 
The results show that the Galileon modifies substantially the dynamics of the collapse, thus making it possible to 
observationally constrain the parameters of this theory.

\end{abstract}

\noindent{\it Keywords}: Modified gravity, Galileon, Spherical Collapse, Cosmological Perturbations

\maketitle

\section{Introduction}

The discovery that the Universe underwent a phase of accelerated expansion at late times, through the study of the distance-redshift relation of type-Ia Supernovae (SNIa) 
\cite{Perlmutter1999a,Riess1998a,Riess1999b}, opened a new scenario for theoretical cosmology: the possibility to live in a Universe whose dynamics is presently driven 
by a component responsible for an ``obscure'' repulsive force, which has been dubbed Dark Energy (DE). Such a component 
should fill $74\%$ of the energy budget of the universe, and it can be obtained by either just considering a non-zero cosmological constant term ($\lcdm$ model). This model fits very 
well observational data, but, up to now, it is impossible 
to give a physical meaning to the tiny value of $\Lambda$ required to explain dark energy. Thus, cosmologists explored alternative theories by e.g, modifying the Einstein-Hilbert action:
\beq
S=\frac{\Mpl^2}{2}\int {\rm d}^4 x \sqrt{-g}\,R+\int {\rm d}^4 x\, {\cal L}_{M}\,,
\eeq
where $\Mpl$ represents the reduced Planck mass. Models that have been proposed are scalar-tensor theories \cite{Brans1961}, $f(R)$ gravity (for a review see \cite{DeFelice2010b}), 
massive gravity (see \cite{Hinterbichler2011d}), Brane-World models (e.g. \cite{Maartens2005}) and others.

Recently, a new class of theories was introduced by \textit{Nicolis et al.} \cite{Nicolis2009}, the so-called {\it Galileon} \cite{Khoury2004b,kobayashi2010cosmic,VanAcoleyen2011,Goon2011c,Qiu2011e,Gao2011,Burrage2011a,Khoury2011a,Trodden2011a,DeFelice2011b,Babichev2011,Padilla2010,Kobayashi2010,Silva2009,Rham2010a,Deffayet2010,Easson2011}. This model was constructed as an effective field 
theory, which is based upon and aims at extending the Dvali-Gabadadze-Porratti model (DGP) \cite{Dvali2000d}. It is interesting because it solves ghost instabilities, which plague DGP, and has a screening mechanism that allows to satisfy the bounds coming from solar system experiments. To avoid the appearance of ghosts it is important to keep the equation of motion up to second-order in time-derivatives. Unfortunately, in the original model, this property was respected only in flat space-time. The works by \textit{Deffayet et al.} \cite{Deffayet2009,Deffayet2009c} found a way to generalize Galileons to curved space-time. To do this, it is necessary to add some extra terms which couple the scalar field with curvature terms. The result is a scalar-tensor theory in which the action, in flat space-time, is invariant under Galilean symmetry ($\partial_\mu\phi\rightarrow\partial_\mu\phi+b_\mu$).

Even though in this paper we study the effects of the late-time cosmic acceleration produced by the scalar field, adopting the spherical collapse model, it is worth mentioning that the importance of the Galileon field also relis on the fact that it can inspire some ``inflationary-like'' model, e.g. \cite{Creminelli2010a} (even though this theory does not respect the galilean symmetry).

In this paper we will use the background evolution given by the tracker solution found in \cite{DeFelice2010h}, which ensures a de Sitter (dS) stable point. While it is shown that the cosmology of the Galileon lets the universe expands accelerating at late-times, in this paper we will show that at short distances we can satisfy solar system constraints via the Vainshtein mechanism \cite{Vainshtein1972,Kaloper2011a}, see also \cite{Kimura2012} for a discussion in the most general second order scalar-tensor theory. This mechanism  can work in massive gravity, but also in different contexts. An example is DGP theory, which possesses a Vainshtein radius defined by $r_V=(r_s r_c^2)^{1/3}$ (where $r_s$ is the Schwarzschild radius of the source, $r_c$ is a coupling constant which defines the crossover scale between a 5-dimensional Minkowsky space and the embedded 4-dimensional space-time). Even if the Galileon does not already have a well-defined Vainshtein radius, we will show how to recover a valid definition of it. In fact, as in DGP, instead of a massive graviton, this mechanism can also work using non-linear self-interaction terms of the scalar field (as $\Box\phi{\left(\nabla\phi\right)}^2$).

The spherical collapse model (e.g. \cite{Schmidt2009f,Schmidt2010,Kimura2011}) studies the evolution of a spherical Dark Matter (DM) overdensity to explain the formation of cosmic structures. We will use the 
top-hat approximation, taking into account the energy non-conservation problem noted in \cite{Schmidt2010}. This problem affects theories with a time-dependent dark energy component, and it can substantially modify the 
virialisation process.

The paper is organized as follows. In Section \ref{SEC:Action} we define the action we are assuming and we obtain the equations of motion. 
In Section \ref{SEC:Background} we briefly review the background evolution of a Friedmann-Lema\^{i}tre-Robertson-Walker universe (FRLW) following a tracker solution found in \cite{DeFelice2010h}. 
In Section \ref{SEC:Perturbations} we study scalar perturbations, both in the linear and non-linear regime. We also study the Vainshtein mechanism, 
and discuss the existence of a solution for the Galileon field in the non-linear regime. In Section \ref{SEC:SphericalCollapse} we study the dynamics of a spherical top-hat matter perturbation. 
In Section \ref{SEC:Conclusions} we discuss our main results. In \ref{Appendix:EOMComponents} and \ref{Appendix:BackgroundFunctions} we give some useful functions.

Throughout the paper we adopt units $c=\hbar=G=1$, except where explicitly indicated; our signature is $(-,+,+,+)$.

\section{Action and Field equations}\label{SEC:Action}

Let us start with the covariant action for the Galileon model non-minimally coupled to the metric \cite{DeFelice2010h}:
\beq \label{EQ:Action}
S=\int {\rm d}^4 x \sqrt{-g}\,\left[ \frac{\Mpl^2}{2}R+\frac{1}{2} \sum_{i=1}^5 c_i {\cal L}_i \right]+\int {\rm d}^4 x\, {\cal L}_{M}\,,
\eeq
where $c_i$ are dimensionless constants. We consider ${\cal L}_{M}$ as the Lagrangian of a pressurless perfect fluid with density $\rho$. The five Lagrangian densities for the scalar field are:
\beqra
{\cal L}_1&=&M^3 \phi \\
{\cal L}_2&=&(\nabla \phi)^2 \\
{\cal L}_3&=&(\square \phi) (\nabla \phi)^2/M^3 \\
{\cal L}_4&=&(\nabla \phi)^2 \left[2 (\square \phi)^2-2 \phi_{;\mu \nu} \phi^{;\mu \nu}-R(\nabla \phi)^2/2 \right]/M^6 \label{EQ:L4}\\
{\cal L}_5&=&(\nabla \phi)^2 [ (\square \phi)^3-3(\square \phi)\,\phi_{; \mu \nu} \phi^{;\mu \nu}+2{\phi_{;\mu}}^{\nu} {\phi_{;\nu}}^{\rho}
{\phi_{;\rho}}^{\mu}+\nonumber\\
&&-6 \phi_{;\mu} \phi^{;\mu \nu}\phi^{;\rho}G_{\nu \rho}] /M^9\label{EQ:L5}\,,
\eeqra
where $M$ is a constant with dimensions of mass, and we defined its value as $M^3\equiv\Mpl H_{dS}^2$. $\Hds$ is the value of the Hubble parameter $H(t)$ in a FRLW 
universe at the de Sitter fixed point. Indeed, as we will see, \cite{DeFelice2010h} found a tracker solution that ends at a stable point called ``de Sitter point'', at which the 
energy density of the scalar field dominates. ${\cal L}_1$ can be understood as a potential term 
and for this reason we set $c_1=0$, since we are interested in analyzing the contribution of the new kinetic terms (the case in which a standard minimally coupled scalar field is 
introduced in the field equations was already studied in \cite{Creminelli2010}). Moreover, with this choice we can employ the tracker solution given in \cite{DeFelice2010h}, that is not admitted if $c_1\neq 0$.
${\cal L}_2$ is the standard kinetic term. ${\cal L}_3$ comes directly from the decoupling limit of DGP theory. ${\cal L}_4$ and ${\cal L}_5$ provide the full generalization 
of an action containing at most second derivatives with respect to Galilean shift symmetry in a flat space-time. The coupling between $\phi$ and the curvature tensors 
are required to construct a Lagrangian free of third or higher-order derivatives in the equations of motion.

Varying this action with respect to the metric $g_{\mu \nu}$ and the scalar field $\phi$ we obtain the equations of motion. For the metric:
\beq \label{EQ:Einstein}
G_{\mu\nu}=\Mpl^{-2}\left[T^{(m)}_{\phantom{(m)}\mu\nu}+T^{(\phi)}_{\phantom{(\phi)}\mu\nu}\right] \;,
\eeq
where
\beq
T^{(\phi)}_{\phantom{(m)}\mu\nu}=\sum_{i=1}^5 c_i T^{(i)}_{\phantom{(m)}\mu\nu} \;,
\eeq
the terms $T^{(i)}_{\phantom{(i)}\mu\nu}$ being listed in \ref{Appendix:EOMComponents}. Instead, varying with respect to the scalar field, we obtain
\beq \label{EQ:Galileon}
\sum_{i=1}^5 c_i \xi^{(i)}=0 \;,
\eeq
where $\xi^{(i)}$ are also listed in \ref{Appendix:EOMComponents}.

\section{Background evolution}\label{SEC:Background}

From Eqs. (\ref{EQ:Einstein}) and (\ref{EQ:Galileon}) we can study the background evolution in an expanding FLRW universe with scale factor $a(t)$. Calling $\phi\equiv\phi(t)$ and $\rho\equiv\rho_m(t)+\rho_r(t)$, 
the background scalar field and background matter and radiation density respectively, the field equations read
\beqra
&&3\Mpl^2 H^2=\rho_{\rm \phi}+\rho_m+\rho_r\,,\label{EQ:Friedmann1}\\
&&3\Mpl^2 H^2+2\Mpl^2 \dot{H}=-P_{\rm \phi}
-\rho_r/3\,,\label{EQ:Friedmann2}
\eeqra
and
\beqra
&&c_2 \left[3 H \dphi+\ddphi\right]-\frac{3c_3}{M^3}\dphi \left[3 H^2 \dphi+\dot{H} \dphi+2 H \ddphi\right]+\frac{18c_4}{M^6}H\dphi^2 \left[3 H^2 \dphi+\right.\label{EQ:BackgroundGalileon}\\
&&\left.+2\dot{H} \dphi+3 H\ddphi\right]-\frac{15c_5}{M^9}H^2\dphi^3 \left[3 H^2 \dphi+3 \dot{H} \dphi+4 H \ddphi\right]=0 \;, \nonumber
\eeqra
where
\beqra
\rho_{\rm \phi} &\equiv& -\frac{c_2}{2}\dphi^2+\frac{3c_3}{M^3}H \dphi^3- \frac{45 c_4}{2M^6} H^2 \dphi^4+\frac{21c_5}{M^9}H^3\dphi^5\,,\\
P_{\rm \phi} &\equiv& -\frac{c_2}{2}\dphi^2-\frac{c_3}{M^3}\dphi^2 \ddphi+\frac{3c_4}{2 M^6}\dphi^3 [8H\ddphi +(3H^2+2\dot{H})\dphi]+\nonumber\\
&&-\frac{3c_5}{M^9}H \dphi^4 [5H \ddphi+2(H^2+\dot{H})\dphi]\,,
\eeqra
are scalar field density and pressure, respectively.

As in \cite{DeFelice2010h}, to study the background we work with the new variables
\beq
r_1 \equiv \dphids\Hds/(\dphi H)\,,
\quad
r_2 \equiv (\dphi/\dphids)^4/r_1\,,
\quad
\Omega_r=\rho_r/(3\Mpl^2 H^2)\,,
\eeq
where $\dphids$ is the time derivative of the scalar field at the dS point. At this point Eqs. (\ref{EQ:Friedmann1}) and (\ref{EQ:Friedmann2}) becomes:
\beqra
&&c_2 \xds^2=6+9\a-12\b\,,\\
&&c_3 \xds^3=2+9\a-9\b\,,
\eeqra
where $\xds \equiv \dphids/(\Hds\Mpl)$. These equations give two conditions for the coefficients $c_2$ and $c_3$. We also set $\a \equiv c_4 \xds^4$ and $\b \equiv c_5\xds^5$; 
therefore our free parameters become $\a$, $\b$ and $\xds$. For simplicity, the assumption $\xds= 1$ will be often used in the 
rest of the paper. An approximation we have done is $\Hds\simeq H_0$, where $H_0$ is the value of the Hubble parameter today.

\begin{figure}[tb]
 \begin{center}
  \includegraphics[width=0.6\textwidth]{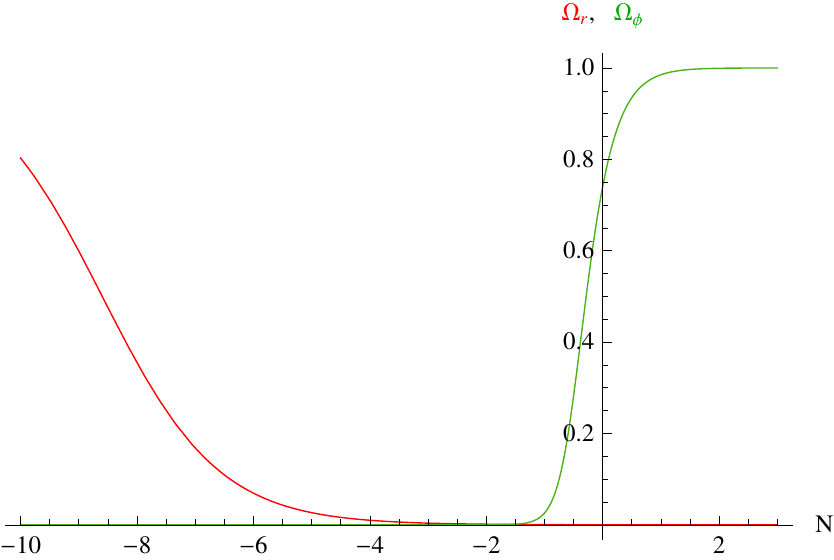}
 \end{center}
 \caption{\textit{In the figure we show the evolution of $\Omega_r$ (red line) and $\Omega_\phi$ (green line), functions of $N=\ln a$.}}\label{FIG:BackgroundEvolution}
 \end{figure}

As we already mentioned, \cite{DeFelice2010h} found a stable tracker solution ($r_1=1$), which drives the universe expansion from the radiation-dominated epoch ($r_2\ll 1$, $\Omega_r=1$), 
through the matter-dominated epoch ($r_2=1$, $\Omega_r\ll 1$), until the dS point ($r_2=1$, $\Omega_r=0$). Note that along $r_1=1$, $\Omega_\phi\equiv\rho_\phi/(3 \Mpl^2 H^2)=r_2$. 
Following this solution, Eqs. (\ref{EQ:Friedmann2}) and (\ref{EQ:Galileon}) with our new variables can be written as
\beq \label{EQ:EvolutionR2}
r_2' =\frac{2 r_2 \left( 3-3r_2+\Omega _r\right)}
{1+r_2},\qquad
\Omega_r' =\frac{\Omega _r
\left(\Omega _r-1-7 r_2\right)}{1+r_2} \;,
\eeq
where primes denote differentiation w.r.t. $N=\ln a$. In Fig. \ref{FIG:BackgroundEvolution} we show the numerical solution of these equations with boundary conditions 
$\Omega_{r_0}=4.8\cdot 10^{-5}$ and $\Omega_{\Lambda_0}=0.74$, where $\Omega_{r_0}$ and $\Omega_{\Lambda_0}$ are the density parameter values today, for 
the radiation and the dark energy component,  respectively. 
These equations cannot be solved analytically; however we have found two analytic functions that approximate the numerical results with an accuracy better than $1.2\%$ at redshift $z\lesssim 21$:
\beq
r_2(N)\simeq 1+\left[\frac{{(1-\Omega_{\Lambda_0})}^2}{2\Omega_{\Lambda_0}}-\frac{1-\Omega_{\Lambda_0}}{2\sqrt{\Omega_{\Lambda_0}}}\cdot\sqrt{4 e^{6 N} + 
\frac{{(1-\Omega_{\Lambda_0})}^2}{\Omega_{\Lambda_0}}}\right]\cdot e^{-6 N}\,,
\eeq
and:
\beq
\Omega_r(N)\simeq 2\Omega_{r_0} e^{-N} {\left(1-\Omega_{\Lambda_0}+\sqrt{4\Omega_{\Lambda_0} e^{6 N}+{(1-\Omega_{\Lambda_0})}^2}\right)}^{-1}\,.
\eeq

To study the stability of the solution $r_1(N)=1$, Eqs. (\ref{EQ:Friedmann1}), (\ref{EQ:Friedmann2}) and (\ref{EQ:BackgroundGalileon}) can be expanded  at linear order in perturbations 
$\delta r_1$, $\delta r_2$ and $\delta \Omega_r$. Thus, it can be obtained:
\beq
{\delta r_1}'(N)=-\frac{9+\Omega_r(N)+3 r_2(N)}{2\left(1+r_2(N)\right)}\delta r_1(N)\,,
\eeq
which reads:
\beq
\delta r_1(N)=\delta r_1(0) \exp\left[-\int_0^N dN' \frac{9+\Omega_r(N')+3 r_2(N')}{2\left(1+r_2(N')\right)}\right]\leq f_0\, e^{-\frac{9}{2}N}\,.
\eeq
$f_0$ is a finite integration constant, and this relation proves that any solution that approaches $r_1(N)=1$, finally reaches it. Indeed, in the rest of the paper, we shall suppose that 
at least after the matter-dominated epoch the evolution of the universe can be described by $\delta r_1\ll 1$.

In \cite{DeFelice2010h}, the authors also find constraints on the parameters $\a$ and $\b$ (assuming $\xds=1$). These constraints follow from the requirement of ghost avoidance. 
They study scalar (S) and tensor (T) perturbations, expanding the action Eq. (\ref{EQ:Action}) at second-order in perturbation theory (see \cite{DeFelice2009c,DeFelice2010f}, for the complete procedure), 
finding conditions for the sign of the kinetic term ($Q_S$ and $Q_T$) and the squared sound speed ($c_S^{\phantom{S}2}$ and $c_T^{\phantom{T}2}$). Thus, in every epoch we have four conditions 
that must be satisfied. Reminding that $\a$ and $\b$ are constants, we can find a region of parameter space where no ghost modes exist. This area is bounded by the analytic functions
 \beq\label{EQ:NoGhostCondition}
    \left\{\begin{array}{l}
          \a>2\, \b\\
 	 \a<2\, \b+2/3\\
 	 \a<12\,\sqrt{\b}-9\,\b-2\\
 	 \a>12/13\, \b+10/13\,.
         \end{array}\right.
\eeq

\section{Cosmological perturbations}\label{SEC:Perturbations}

In this section we study the evolution of scalar perturbations on sub-horizon scales. Our work focuses on the dynamics of a spherically symmetric perturbed metric. Let us choose
the conformal Newtonian gauge,
\beq
ds^{2}=-(1+2\Psi)dt^{2}+a^{2}(t)(1+2\Phi)\delta_{ij}dx^{i}dx^{j}\,.
\eeq
Perturbations of the energy density and the scalar field are given by
\beq
\rho(\vec{x},t)\equiv \rho_0 (t)+\delta\rho (\vec{x},t)\qquad \phi(\vec{x},t)\equiv\phi_0(t)+\varphi(\vec{x},t) \,.
\eeq
In the following we will drop the suffix ``$0$''. In this regime there are two valid approximations that simplify the field equations. 
The first one is the sub-horizon approximation $\mathcal{O}(\nabla^2\Phi/a^2) \gg \mathcal{O}(H^2\Phi)$. The second one is the quasi-static approximation,  which allows us to neglect time derivatives of perturbations compared with space derivatives, assuming we are working with non-relativistic matter at short distances.

\subsection{Linear perturbation theory}\label{SEC:LinearTheory}

Replacing physical gradients with comoving gradients, at linear order Eqs. (\ref{EQ:Einstein}) and (\ref{EQ:Galileon}) become ($\nabla$ denotes a spatial gradient):
\beq\label{EQ:LinearEinstein00}
\left(2 \Mpl^2+\dphi^2\g_1(t)\right)\nabla^2\Phi=-\delta\rho+\g_2(t)\nabla^2\varphi\,,
\eeq
\beq\label{EQ:LinearEinsteinij}
\left(2 \Mpl^2+3\g_3(t)\right) \nabla^2\Phi+\left(2 \Mpl^2+\dphi^2\g_1(t)\right)\nabla^2\Psi=3\g_4(t)\nabla^2\varphi\,,
\eeq
and
\beq\label{EQ:LinearGalileon}
\g_5(t)\nabla^2\varphi+\g_2(t)\nabla^2\Psi+3\g_4(t)\nabla^2\Phi=0\;, 
\eeq
where $\g_i(t)$ are functions of the background, whose explicit form is given in \ref{Appendix:BackgroundFunctions}.

It is important to note that one of the differences between these equations and those for the kinetic braiding model studied in \cite{Kimura2011} is 
the presence of an anisotropic stress in the RHS of Eq. (\ref{EQ:LinearEinsteinij}).

Manipulating Eqs. (\ref{EQ:LinearEinstein00}) and (\ref{EQ:LinearEinsteinij}), we obtain the modified Poisson equation
\beq \label{EQ:PoissonEquation}
\frac{{\left(2 \Mpl^2+\dphi^2\g_1\right)}^2}{2 \Mpl^2+3\g_3}\nabla^2\Psi=\delta\rho-\left[\g_2-3\g_5\frac{2 \Mpl^2+\dphi^2\g_1}{2 \Mpl^2+3\g_3}\right]\nabla^2\varphi\,.
\eeq

Using Eqs. (\ref{EQ:LinearGalileon}), (\ref{EQ:LinearEinstein00}) and (\ref{EQ:PoissonEquation}), the differential equation for the evolution of the scalar field takes the form
\beq\label{EQ:LinearPoissonGalileon}
\nabla^2\varphi=A(t)\,\delta\rho(t,\vec{r}),
\eeq
where
\beq
A(t)\equiv\frac{\g_2(t)\g_7(t)-3\g_4(t)\g_6(t)}{{\g_2(t)}^2\g_7(t)-\g_5(t){\g_6(t)}^2-6\g_2(t)\g_4(t)\g_6(t)} \;,
\eeq
with $\g_6(t)\equiv\left(2 \Mpl^2+\dphi^2\g_1(t)\right)$ and $\g_7(t)\equiv\left(2 \Mpl^2+3\g_3(t)\right)$. Considering a spherically symmetric object of radius $R_S$, 
we can easily integrate Eq. (\ref{EQ:LinearPoissonGalileon}) to obtain an analytic expression for the evolution of the scalar field. 
Defining $m(t,r)\equiv4\pi\int_0^r\rm d r' {r'}^2 \delta\rho$, we obtain
\beq\label{EQ:LinearSolutionGalileon}
\frac{\rm d \varphi}{\rm d r}=\frac{A(t)\,m(t,r)}{4\pi r^2}+\frac{C}{r^2}\,,
\eeq
where $C$ is an integration constant that, outside the source, can be viewed as an increase in $M_s\equiv m(t,R_S)$. While this term is present in $\varphi'$, it does not enter in $\nabla^2\varphi$, so that 
the gravitational potential is not affected by our choice of $C$. Therefore, for our purposes we can set $C=0$.

\subsection{Vainshtein mechanism in the linear regime}

The Vainshtein mechanism works by screening the effects of the scalar field on the gravitational potential at small distances, so that one can satisfy the constraints coming from solar-system tests, while preserving
the accelerated expansion of the universe on cosmological scales. The difference between this mechanism and the Chamaleon one is that the first also works by using non-linearities of the perturbations to this aim. 
At large distances ($r\gg r_V$, where $r_V$ is the Vainshtein radius of the source) linear terms of the scalar field become dominant, while for $r\ll r_V$ non-linear terms become dominant 
(these terms will be shown in Eqs. (\ref{EQ:NonLinearEinstein00}), (\ref{EQ:NonLinearEinsteinij}) and (\ref{EQ:NonLinearGalileon})). 
This is called ``self-screening effect''. A discussion about the magnitude of the Vainshtein radius ($r_V$) of a spherically symmetric source will be given later (Sec. \ref{SEC:VainshteinRadius}).

A first approach is to study within the linear approximation the contribution of the scalar field to the gravitational potential. Recalling Eq. (\ref{EQ:PoissonEquation}), to have a qualitative knowledge 
that outside the Vainshtein radius the scalar field drives the late time cosmic acceleration, we have to compare the contribution of the gravitational with the scalar field 
intensity \cite{Burrage2010}. Indeed, our request is that the two are comparable:
\beq\label{EQ:GalileonContribution}
\frac{\varphi'(r)}{\Psi'(r)}\simeq 1\,.
\eeq

It can be shown that the above ratio is a monotone function, which starts from $\simeq0$ during the radiation-matter-dominated epoch. At the dS point, recalling 
Eq. (\ref{EQ:LinearPoissonGalileon}) with $\xds=1$, we obtain
\beq
{\left|\frac{\varphi'(r)}{\Psi'(r)}\right|}_{\rm dS}=\left|\frac{A(t_{\rm dS})}{4 \pi}\right| =\left|\frac{1}{24\pi \Mpl (2 \b -\a)}\right|\,.
\eeq
Taking into account the region in the plane $(\xds=1,\b,\a)$ bounded by the no-ghost condition (\ref{EQ:NoGhostCondition}), it can be shown that the magnitude of the last ratio at the dS point is bounded by
\beq\label{EQ:A}
\frac{1}{4\sqrt{2\pi}}<{\left|\frac{\varphi'(r)}{\Psi'(r)}\right|}_{\rm dS}<+\infty
\eeq
This result means that the contribution of the scalar field at the dS point on scales $r\gg r_V$ is always important, and the importance can be set choosing 
proper values for $\a$ and $\b$. In particular we can find a couple $(\a,\b)$ which satisfies Eq. (\ref{EQ:GalileonContribution}).

\begin{figure}[htb]
\begin{center}
 \includegraphics[width=0.8\textwidth]{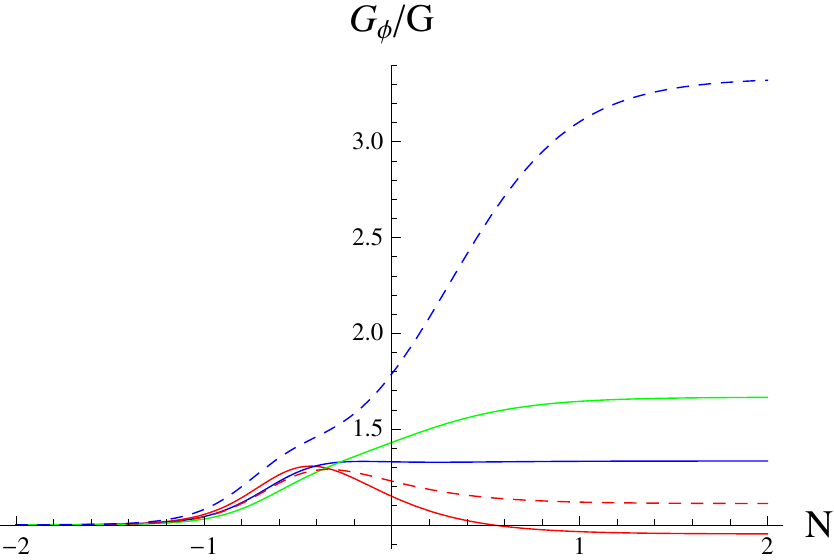}
\caption{\textit{This plot shows the evolution of $G_{\phi}$, with $\xds=1$, in different cases.The values for $(\a,\b)$ are: $(-1, -0.55)$, blue dashed line; $(-0.45, -0.4)$, 
red line; $(-0.2, -0.2)$, green line; $(-0.55, -0.4)$, blue solid line; $(0.1, -0.1)$, 
red dashed line.}}\label{FIG:Geffx1}
\end{center}
\end{figure}

\begin{figure}[htb]
\begin{center}
 \includegraphics[width=0.8\textwidth]{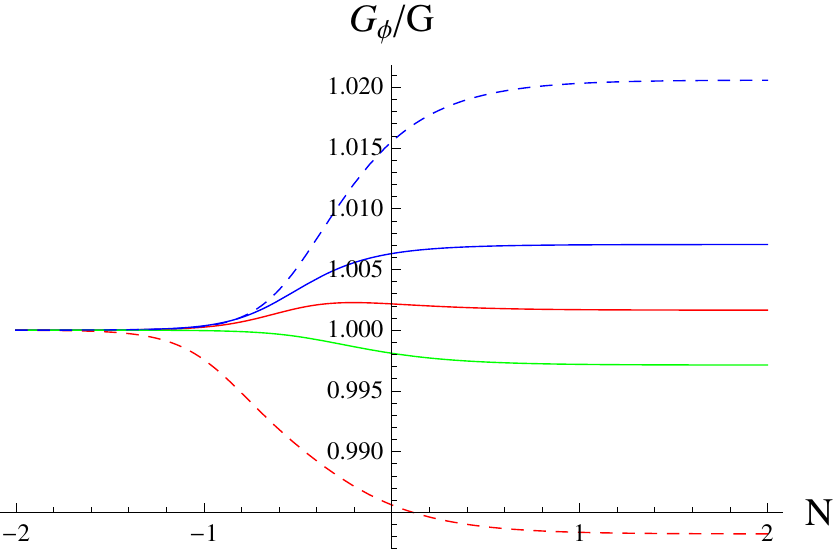}
\caption{\textit{The same as in Fig. \ref{FIG:Geffx1}, but with $\xds=0.3$.}}\label{FIG:Geffx03}
\end{center}
\end{figure}

\begin{figure}[htb]
\begin{center}
 \includegraphics[width=0.8\textwidth]{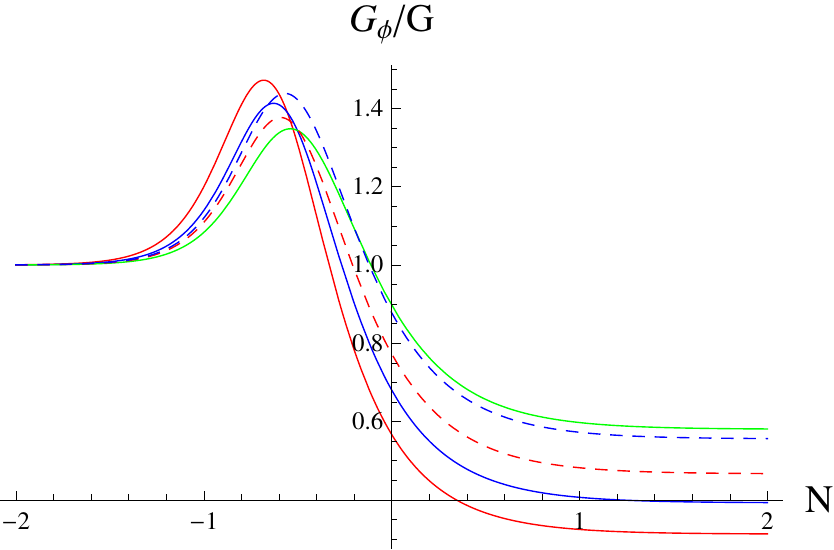}
\end{center}
\caption{\textit{The same as in Fig. \ref{FIG:Geffx1}, but with $\xds=1.2$.}}\label{FIG:Geffx12}
\end{figure}

With Eq. (\ref{EQ:LinearPoissonGalileon}) we can write the modified Poisson equation (\ref{EQ:PoissonEquation}) in a more convenient form:

\beq
\nabla^2\Psi=4 \pi G_{\phi}\delta\rho(t,\vec{r})\,,
\eeq

where:
\beq
G_{\phi}(t)=\frac{\g_5(t)\g_7(t)+9{\g_4(t)}^2}{4\pi\left[6\g_2(t)\g_4(t)\g_6(t)-{\g_2(t)}^2\g_7(t)+\g_5(t){\g_6(t)}^2\right]}\,.
\eeq
The modified gravitational constant assumes the value of the Newtonian one during the radiation-matter-dominated era, while it is 
\beq
G_{\phi}(t_{\rm dS})=\frac{G}{3 (\a-2\b)}
\eeq
at the dS point (when $\xds=1$). The limit $\xds\rightarrow0$ gives us the usual GR result $G_{\phi}(t_{\rm dS})=G$. Instead, the limit 
$\xds\rightarrow\infty$ gives $G_{\phi}(t_{\rm dS})\rightarrow0$, which means, as expected, that the effective gravitational constant becomes 
small w.r.t. the Newtonian one ($G\varpropto \Mpl^{-2}$). The plots in Figs. \ref{FIG:Geffx1}, \ref{FIG:Geffx03} and \ref{FIG:Geffx12} show 
that we can vary the asymptotic value of $G_{\phi}$ as we desire, to obtain, in principle, any reasonable model for the late time cosmic acceleration. 
The difference between the three graphs is the value of the parameter $\xds$, which sets the contribution of the Galileon field at the dS point. 
This result also agrees with the expectations of Eq. (\ref{EQ:A}), quantifying the effective contribution of the scalar field at large distances on observables quantities. 
Of course, these results do not represent any realistic model, we are only interested here in investigating the range of possibilities offered by the Galileon theory. Moreover, astrophysical and cosmological constraints on the Galileon model have just started being considered \cite{Brax2011h,DeFelice2011d,Hirano2010e,Hirano2011d}.

\subsection{Non-linear evolution}

When perturbations grow, Eqs. (\ref{EQ:LinearEinstein00}), (\ref{EQ:LinearEinsteinij}) and (\ref{EQ:LinearGalileon}) must be replaced by fully non-linear ones.
Neglecting time-derivatives of perturbations and assuming that the characteristic scale of the perturbation is well within the Hubble radius, we obtain
\beqra\label{EQ:NonLinearEinstein00}
&&\left(2 \Mpl^2+\dphi^2\g_1(t)\right)\nabla^2\Phi=-\delta\rho+\g_2(t)\nabla^2\varphi +\g_1(t)\left[{(\nabla^2\varphi)}^2+\right.\nonumber\\
&&\left.-\nabla^{i}_{\phantom{i}j}\varphi\nabla^{j}_{\phantom{j}i}\varphi\right]+\e_1(t)\left[{(\nabla^2\varphi)}^3+2\nabla^{i}_{\phantom{i}j}\varphi\nabla^{j}_{\phantom{j}k}\varphi\nabla^{k}_{\phantom{k}i}\varphi+\right.\nonumber\\
&&\left.-3\nabla^2\varphi\nabla^{i}_{\phantom{i}j}\varphi\nabla^{j}_{\phantom{j}i}\varphi\right]-\frac{3}{2}\dphi^2\e_1(t)\left[\nabla^2\varphi\nabla^2\Phi-\nabla^{i}_{\phantom{i}j}\Phi\nabla^{j}_{\phantom{j}i}\varphi\right]\,,
\eeqra

\beqra\label{EQ:NonLinearEinsteinij}
&&\left(2 \Mpl^2+3\g_3(t)\right) \nabla^2\Phi+\left(2 \Mpl^2+\dphi^2\g_1(t)\right)\nabla^2\Psi=3\g_4(t)\nabla^2\varphi+\nonumber\\
&&+3\e_2(t)\left[{(\nabla^2\varphi)}^2-\nabla^{i}_{\phantom{i}j}\varphi\nabla^{j}_{\phantom{j}i}\varphi\right]-\frac{3}{2}\dphi^2\e_1(t)\left[\nabla^2\varphi\nabla^2\Psi-\nabla^{i}_{\phantom{i}j}\Psi\nabla^{j}_{\phantom{j}i}\varphi\right]\,.
\eeqra

Eq. (\ref{EQ:LinearGalileon}), instead, takes the form
\beqra\label{EQ:NonLinearGalileon}
&&\g_5(t)\nabla^2\varphi+\g_2(t)\nabla^2\Psi+3\g_4(t)\nabla^2\Phi+\e_3(t)\left[{(\nabla^2\varphi)}^2-\nabla^{i}_{\phantom{i}j}\varphi\nabla^{j}_{\phantom{j}i}\varphi\right]-\nonumber\\
&&-\e_4(t)\left[{(\nabla^2\varphi)}^3+2\nabla^{i}_{\phantom{i}j}\varphi\nabla^{j}_{\phantom{j}k}\varphi\nabla^{k}_{\phantom{k}i}\varphi-3\nabla^2\varphi\nabla^{i}_{\phantom{i}j}\varphi\nabla^{j}_{\phantom{j}i}\varphi\right]+\nonumber\\
&&+2\g_1(t)\left[\nabla^2\varphi\nabla^2\Psi-\nabla^{i}_{\phantom{i}j}\Psi\nabla^{j}_{\phantom{j}i}\varphi\right]+6\e_2(t)\left[\nabla^2\varphi\nabla^2\Phi-\nabla^{i}_{\phantom{i}j}\Phi\nabla^{j}_{\phantom{j}i}\varphi\right]-\nonumber\\
&&-\frac{3}{2}\dphi^2\e_1(t)\left[\nabla^2\Psi\nabla^2\Phi-\nabla^{i}_{\phantom{i}j}\Phi\nabla^{j}_{\phantom{j}i}\Psi\right]+3\e_1(t)\left[{(\nabla^2\varphi)}^2\nabla^2\Psi-\right.\nonumber\\
&&\left.-2\nabla^2\varphi\nabla^{i}_{\phantom{i}j}\varphi\nabla^{j}_{\phantom{j}i}\Psi-\nabla^2\Psi\nabla^{i}_{\phantom{i}j}\varphi\nabla^{j}_{\phantom{j}i}\varphi+2\nabla^{i}_{\phantom{i}j}\varphi\nabla^{j}_{\phantom{j}k}\varphi\nabla^{k}_{\phantom{k}i}\Psi\right]=0\,,
\eeqra
where the $\e_i(t)$ functions are listed in \ref{Appendix:BackgroundFunctions}.

Eqs. (\ref{EQ:NonLinearEinstein00}), (\ref{EQ:NonLinearEinsteinij}) and (\ref{EQ:NonLinearGalileon}) are more complicated than in the linear case, however, assuming spherical symmetry, they are in fact integrable. 
The boundary values of the perturbations can be determined by resorting to the physical meaning to these fields. For example, from GR we know that the physical solution of the Poisson equation is
\beq
{\Psi_{GR}}'(t,r)=\frac{G m(t,r)}{r^2}.
\eeq
Recalling the definition of the mass function, $m(t,r)\equiv4\pi\int_0^r{\rm d}r'r'^2\delta\rho(t,r)$, if there are no singularities at $r=0$ for the density perturbation, 
this relation tells us that ${\Psi_{GR}}'(t,0)=0$ (to violate this limit we have to choose $\delta\rho(r)\propto r^{-n}$, with $n\geq3$). At small scales we want to recover GR, 
so the physical meaning of $\Psi(t,r)$ should be that of gravitational potential ($\Psi'(t,r)\simeq{\Psi_{GR}}'(t,r)$). Indeed, the natural assignment is $\Psi'(t,0)=0$. The same 
argument applies to $\Phi'(t,r\rightarrow 0)\simeq-{\Psi_{GR}}'(t,r\rightarrow 0)$. Instead, the scalar field and its perturbations are not directly observable quantities, 
so we have to choose the correct boundary value by mathematical arguments or by its effect on measurable physical quantities. 
Like in Eq. (\ref{EQ:LinearSolutionGalileon}), at $r\rightarrow 0$ there should be some divergent term. However, the same reasoning used in the 
linear case allows us to consider $\varphi'(r\rightarrow 0)$ finite.

Integrating Eqs. (\ref{EQ:NonLinearEinstein00}), (\ref{EQ:NonLinearEinsteinij}) and (\ref{EQ:NonLinearGalileon}) for a spherically symmetric object, we obtain
\beq\label{EQ:IntegralEinstein00}
\g_6\frac{\Phi'}{r}=-\frac{m(t,r)}{4\pi r^3}+\g_2\frac{\varphi'}{r}+2\g_1\frac{\varphi'^2}{r^2}+2\e_1\frac{\varphi'^3}{r^3}-2\e_1\frac{{\varphi'(0)}^3}{r^3}-3\dphi^2\e_1\frac{\varphi' \Phi'}{r^2}
\eeq
\beq\label{EQ:IntegralEinsteinij}
\g_7\frac{\Phi'}{r}+\g_6\frac{\Psi'}{r}=3\g_4\frac{\varphi'}{r}+6\e_2\frac{\varphi'^2}{r^2}-3\dphi^2\e_1\frac{\varphi'\Psi'}{r^2}
\eeq
\beqra\label{EQ:IntegralGalileon}
\g_5\frac{\varphi'}{r}+\g_2\frac{\Psi'}{r}+3\g_4\frac{\Phi'}{r}+2\e_3\frac{\varphi'^2}{r^2}-2\e_4\frac{\varphi'^3}{r^3}+&&\nonumber\\
+2\e_4\frac{{\varphi'(0)}^3}{r^3}+4\g_1\frac{\varphi'\Psi'}{r^2}+6\e_1\frac{\varphi'^2 \Psi'}{r^3}+12\e_2\frac{\varphi'\Phi'}{r^2}-3\dphi^2\e_1\frac{\Phi'\Psi'}{r^2}&=&0\,,
\eeqra

Note that we have not yet analyzed the case in which the scalar field has a boundary value $\varphi'(r=0)$ finite, but different from zero; to do this we have to impose a physical condition. 
From Eq. (\ref{EQ:IntegralEinsteinij}), we can write
\beqra\label{EQ:BoundaryValueSolutionGalileon}
&&\frac{\varphi'(r)}{r}=-\frac{\g_4}{4 \e_2}+\frac{\dphi^2 \e_1}{4 \e_2}\cdot\frac{\Psi'(r)}{r}+\frac{{\rm Sgn}(\g_4)}{4 \e_2}\left[{\left(-\g_4+\dphi^2\e_1 \frac{\Psi'(r)}{r}\right)}^2+\right.\nonumber\\
&&{\left.+\frac{8}{3}\e_2\left(2 \Mpl^2+3\g_3\right)\frac{\Phi'(r)}{r}+\frac{8}{3}\e_2\left(2 \Mpl^2+\dphi^2\g_1\right)\frac{\Psi'(r)}{r}\right]}^{1/2}\,;
\eeqra
here we have chosen the solution which matches the linear one (\ref{EQ:LinearSolutionGalileon}) when $r\rightarrow\infty$.

Without any loss of generality, the metric perturbations can be written as
\beqra
&&\Psi'(r)=\Psi'_{GR}(r)\left[1+\delta_\Psi(r)\right]=\frac{G\, m(t,r)}{r^2}\left[1+\delta_\Psi(r)\right]\\
&&\Phi'(r)=\Phi'_{GR}(r)\left[1+\delta_\Phi(r)\right]=-\frac{G\, m(t,r)}{r^2}\left[1+\delta_\Phi(r)\right]\,.
\eeqra
In this case, $\Psi_{GR}(r)$ can be understood as the gravitational potential generated by a perturbation in the $\lcdm$ model. 
When $r\ll r_V$, $\delta_\Psi$ and $\delta_\Phi$ have to be small by solar-system constraints ($\delta_\Psi\,,\delta_\Phi\lesssim 10^{-3}$), so we can treat them as 
small perturbations. In this limit, at first order, Eq. (\ref{EQ:BoundaryValueSolutionGalileon}) becomes
\beqra\label{EQ:ApproxSolutionGalileon}
\frac{\varphi'(r)}{r}&\simeq&-\frac{\g_4}{4\e_2}+\frac{\dphi^2\e_1 \Psi'_{GR}(r)}{4\e_2 r}+\frac{{\rm Sgn}(\g_4) f(t,r)}{4\e_2}+\frac{\dphi^2\e_1\Psi'_{GR}(r)}{4\e_2 r}\delta_\Psi(t,r)+\nonumber\\
&&+\frac{{\rm Sgn}(\g_4)\Psi'_{GR}(r)}{12\e_2 f(t,r) r}\left[3\dphi^4{\e_1}^2\frac{\Psi'_{GR}(r)}{r}-3\dphi^2\g_4\e_1+4\g_6\e_2\right]\delta_\Psi(t,r)+\nonumber\\
&&-\frac{{\rm Sgn}(\g_4)\g_7 \Psi'_{GR}(r)}{f(t,r) r}\delta_\Phi(t,r)\,,
\eeqra
where
\beqra
f(t,r)&\equiv&\left[{\g_4}^2+\dphi^4{\e_1}^2\frac{{\Psi'_{GR}(r)}^2}{r^2}-8\g_3\e_2\frac{\Psi'_{GR}(r)}{r}-2\dphi^2\g_4\e_1\frac{\Psi'_{GR}(r)}{r}+\right.\nonumber\\
&&{\left.+\frac{8}{3}\dphi^2\g_1\e_2\frac{\Psi'_{GR}(r)}{r}\right]}^{1/2}\,.
\eeqra

From Eq. (\ref{EQ:ApproxSolutionGalileon}), we are now ready to choose a reasonable boundary value for $\varphi'(r)$. 
It is sufficient to suppose that neither $\delta_\Psi$ nor $\delta_\Phi$ diverge in the limit $r\rightarrow0$, to show that $\varphi'(r\rightarrow0)=0$.

\subsection{Vainshtein radius}\label{SEC:VainshteinRadius}

Having obtained the non-linear equations of motion, we are now ready to investigate the radius at which non-linearities become important. 
The simplest way to estimate $r_V$ is to plug-in the linear solutions into the non-linear equations, and estimate when the non-linear terms 
become comparable with the linear ones. First, considering Eq. (\ref{EQ:IntegralEinstein00}), from the quadratic term we obtain
\beq
\left.2\frac{\g_1}{\g_2}\cdot\frac{\varphi'}{r}\right|_{r=r_{V_1}}\simeq1\,.
\eeq
To solve this equation, we need to know the matter density profile. However, using Eq. (\ref{EQ:LinearSolutionGalileon}), in the general case we find
\beq
{r_{V_1}}^3=\frac{\g_1(t) A(t)}{2\pi\g_2(t)}\left[m(t,r)+\Delta m(t,r,r_{V_1})\right]\,,
\eeq
where $\Delta m(t,r,r_{V_1})=4\pi\int_r^{r_{V_1}} dr' {r'}^2\delta\rho$. The interior solution for a top-hat profile leads to an $r$-invariant equation. 
The simple consideration is that, depending on the epoch and on the choice of the background parameters, we can have this region all inside or all outside 
the Vainshtein region. Instead, outside a source of mass $M_s$ we find (defining $R_V\equiv r_V(R)$)
\beq
{\left(\frac{{R_{V_1}}}{R}\right)}^3=\left|\frac{2\g_1}{\g_2}\cdot\frac{A(t) M_s}{4\pi R^3}\right|\,.
\eeq
The same procedure for the cubic term leads to
\beq
{\left(\frac{R_{V_{2}}}{R}\right)}^3=\left|\sqrt{\frac{2\e_1}{\g_2}}\cdot\frac{A(t) M_s}{4\pi R^3}\right|\,.
\eeq
Comparing the two Vainshtein radii we see that they are comparable. This means that we have an exterior linear region, but, when we enter 
the non-linear one, quadratic and cubic terms can both dominate. Indeed, the contribution derived from the terms $c_4$ and $c_5$ influences in a non-negligible 
way the scalar field profile. This also proves that at sufficiently large distances we recover the predictions of the linear theory, discussed in Sec. \ref{SEC:LinearTheory}.

Other three important Vainshtein radii, coming from Eqs. (\ref{EQ:IntegralEinsteinij}) and (\ref{EQ:IntegralGalileon}), are
\beq
{\left(\frac{{R_{V_3}}}{R}\right)}^3=\frac{2\e_2}{\g_4}\cdot\frac{A(t) M_s}{4\pi R^3}\,,
\eeq
\beq
{\left(\frac{{R_{V_4}}}{R}\right)}^3=\frac{2\e_3}{\g_5}\cdot\frac{A(t) M_s}{4\pi R^3}\,.
\eeq
and:
\beq
{\left(\frac{{R_{V_5}}}{R}\right)}^3=\sqrt{\frac{2\e_4}{\g_5}}\cdot\frac{A(t) M_s}{4\pi R^3}\,.
\eeq

Here we have neglected non-linear interactions which couple $\varphi$ with $\Phi$ and $\Psi$, because they produce results analogous to the previous ones. 
The Vainshtein radius can be set as $R_V\equiv{\rm Max}(R_{V_i})$, where $i=1,..,11$. It is straightforward to prove that $R_V(t\rightarrow-\infty)\rightarrow+\infty$, 
while $R_V(t\rightarrow+\infty)= f(\a,\b,\xds) M_s/(4\pi \Mpl \Hds^2)$, where $f$ is a generic function of the background parameters. 
This result agrees with the predictions of \cite{Schmidt2010} and \cite{Kimura2011}.

\subsection{Galileon field evolution}

In this section we study the Galileon field evolution, starting from Eqs. (\ref{EQ:IntegralEinstein00}), (\ref{EQ:IntegralEinsteinij}) and (\ref{EQ:IntegralGalileon}). 
These are three algebraic equations in $\Psi'(r)$, $\Phi'(r)$ and $\varphi'(r)$, so it is straightforward to obtain a sixth-order polynomial equation in $\varphi'(r)$ 
(to simplify the problem we will work under the assumption that $x_{\rm ds}=1$):
\beqra\label{EQ:NonLinearGalileonEquation}
&&\frac{\varphi'^6}{r^6}+\lambda_1(t)\frac{ \varphi'^5}{r^5}+\lambda_2(t)\frac{\varphi'^4}{r^4}+\left(\lambda_3(t)\delta_m+\lambda_4(t)\right)\frac{\varphi'^3}{r^3}+\left(\lambda_5(t)\delta_m+\right.\nonumber\\
&&\left.+\lambda_6(t)\right)\frac{\varphi'^2}{r^2}+\left(\lambda_7(t)\delta_m+\lambda_8(t)\right)\frac{\varphi'}{r}+\lambda_{9}(t)\delta_m+\lambda_{10}(t){\delta_m}^2=0\,,
\eeqra
where $\lambda_i$ are background functions, combinations of $\g_i$ and $\e_i$. From Eq. (\ref{EQ:NonLinearGalileonEquation}) it follows that $\varphi'(r)$ has six branches of solutions. 
What is the correct one? Remembering the Vainshtein effect, we want that the physical solution reduces to Eq. (\ref{EQ:LinearSolutionGalileon}) at large distances. 
Of course, this condition cannot be verified analytically, but it is sufficient to choose between the real solutions of Eq. (\ref{EQ:NonLinearGalileonEquation}).

Are we sure that, for a given couple $(\a,\b)$, Eq. (\ref{EQ:NonLinearGalileonEquation}) has at least a couple of solutions during the whole evolution of the universe? 
Obviously this condition is not sufficient to ensure the existence of the physical solution, but it is a necessary condition. 
In the linear regime the existence of a physical solution was proved in Sec. \ref{SEC:LinearTheory}, thus the problems can be inside the Vainshtein radius. 
As proved in Sec. \ref{SEC:VainshteinRadius}, at small distances non-linear terms become dominant for the evolution of the scalar field. 
In particular, instead of Eq. (\ref{EQ:NonLinearGalileonEquation}), we can work with the equation
\beq\label{EQ:ApproxNonLinearGalileon}
\frac{\varphi'^6}{r^6}+\lambda_1(t)\frac{ \varphi'^5}{r^5}+\lambda_2(t)\frac{\varphi'^4}{r^4}+\lambda_{10}(t){\delta_m}^2=0.
\eeq
Also in this case we do not have an analytic solution for the scalar field; however Eq. (\ref{EQ:ApproxNonLinearGalileon}) 
gives new constraints on the allowed region in the parameter space $(\a,\b)$.

Consider a function like
\beq\label{EQ:6OrderPolynomial}
f(x)=x^6+A\,x^5+B\,x^4+C\,,
\eeq
where $A,\,B,\,C\neq 0$ are real coefficients. The RHS of this equation has the same form as Eq. (\ref{EQ:ApproxNonLinearGalileon}), after the substitution $\varphi'(t,r)/r\rightarrow x$. 
It was demonstrated that there is no analytic method to find a solution for  $f(x)=0$, when $f(x)$ is a fifth or higher degree polynomial. However, since
\beq
\lim_{x\to \pm \infty} f(x)=+\infty\,,
\eeq
it is sufficient to require that a minimum of this function is $<0$, to be sure to have at least a couple of real solutions. 
The points which satisfy $f'(x)=0$ are
\beq
x_{1,2,3}=0\qquad x_{4,5}=-\frac{5}{12}A \pm \sqrt{\frac{25}{144}A^2-\frac{2}{3}B}\,.
\eeq
The zeros of Eq. (\ref{EQ:NonLinearGalileonEquation}) can be understood as six perturbative terms around $x_i$. 
Let us assume that, for the purpose of this section, these perturbations are small. The set of parameters for which $f(x)=0$ has at least a couple of solutions, which are given by
\beq
f(x_1)<0\quad\vee\quad f(x_4)<0\quad \vee\quad f(x_5)<0\,.
\eeq

\begin{figure}[!tb]
\begin{center}
 \includegraphics[width=0.6\textwidth]{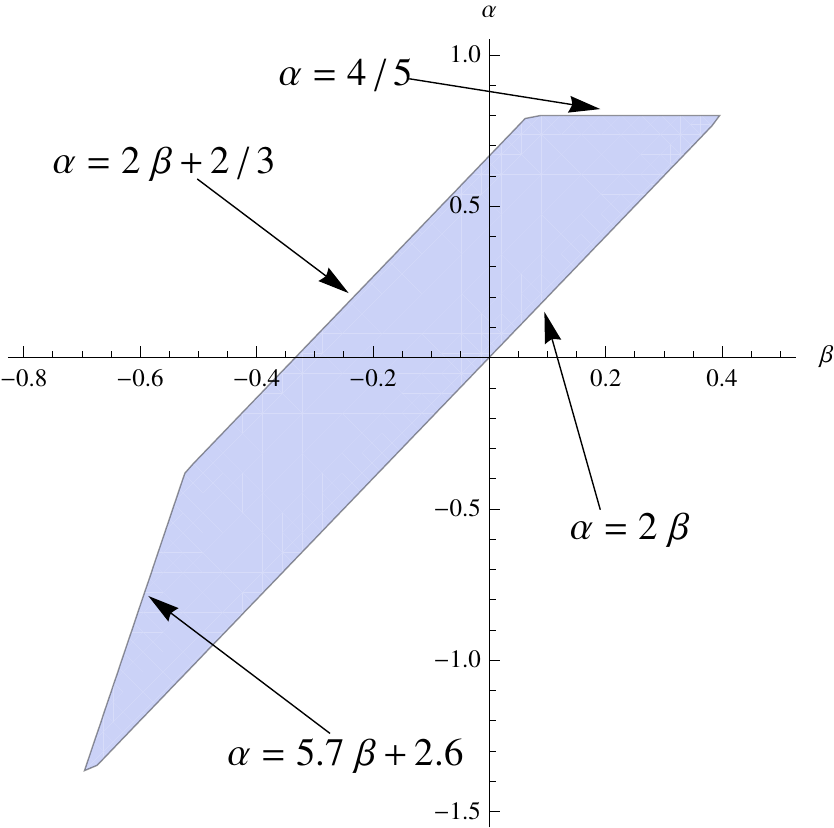}
\end{center}
\caption{\textit{In this figure we show the allowed region in the plane $(\b,\a)$ obtained by mixing the no-ghost conditions and the conditions 
for the existence of the scalar field in the non-linear regime.}}\label{FIG:AllowedParametersRegion}
\end{figure}

Substituting our background functions into the parameters $A,\, B,\, C$, we must pay attention to the dependence on $t$, because the previous inequalities have to be hold true $\forall\, t$. 
The first one becomes
\beq
f(x_1)=\frac{H_{\rm ds}^{12}\, M_{\rm Pl}^6}{144\, \dphi^4\, \b^2}\cdot\frac{\left(\frac{\dphi}{H_{\rm ds} M_{\rm Pl}}\right)^4\left[\a+ 6\b \left(\frac{\ddphi}{H_{\rm ds}^2 M_{\rm Pl}}\right)\right]-2}{4+\left(\frac{\dphi}{H_{\rm ds} M_{\rm Pl}}\right)^4\left[-5\a+42\b \left(\frac{\ddphi}{H_{\rm ds}^2 M_{\rm Pl}}\right)\right]} \delta_m(t)^2<0\,.
\eeq
It can be proved that this condition is verified $\forall\, t$, when
\beq
 \left\{\begin{array}{l}
         \a<4/5\\
	 \a\lesssim5.22 \b+1.93\\
	 \a\lesssim-3.73 \b+4.83\,.
        \end{array}\right.
\eeq
These relations were obtained evaluating the above expression at some critical times, when $f(x_1)$ results maximized/minimized. 
We were able to do this because $f(0)$ takes a simple form, but this is not the case for $f(x_4)$ and $f(x_5)$. In fact, the form of these functions at the points $x_{4,5}$ is
\beqra
f(x_{4,5})&=&C-\frac{2}{12^{6}}{\left(\pm 5 A+\sqrt{25 A^2-96 B}\right)}^4\left(5 A^2-24 B+\right.\nonumber\\
&&\left.\pm A\sqrt{25 A^2-96 B}\right)\,.
\eeqra
In our case, the parameter $C$ depends on the matter-density perturbation, so the inequalities which follow from the above expression have to be evaluated in two distinct cases. 
The first one is when the density term dominates on the other terms (the analysis is the same as in $f(0)<0$ case), the second one when it is subdominant. 
The latter case involves more complicated expressions for the parameters $\a$ and $\b$, so we were only able to solve it numerically. 
Combining these results with the no-ghost condition given in \cite{DeFelice2010h}, the constraints on the parameters $\a$ and $\b$ become
\beq\label{EQ:NewParameterConditions}
 \left\{\begin{array}{l}
         \a>2\, \b\\
	 \a<2\, \b+2/3\\
	 \a<4/5\\
	 \a\lesssim 5.7\, \b+2.62\,,
        \end{array}\right.
\eeq
and are represented in Fig. \ref{FIG:AllowedParametersRegion}.

\section{Spherical Collapse}\label{SEC:SphericalCollapse}

In this section we will restrict our analysis to a top-hat matter configuration
\begin{eqnarray}
  \rho(r)= \left\{ \begin{array}{ll}
    \rho_0+\delta\rho & ~~~~r\leq R \\
    \rho_0 & ~~~~r>R\\
\end{array} \right.,
\hspace{0.1cm}
  m(r)= \left\{ \begin{array}{ll}
    \delta M\, (r/R)^3 & ~~~~r\leq R \\
    \delta M & ~~~~r>R\\
\end{array} \right..
\end{eqnarray}

The mass $\delta M$ is the total mass of the density perturbation $\delta\rho$, while $M\equiv 4/3\, \pi\,(\rho_0+\delta\rho)\,R^3$. The two masses are related by
\beq
\delta M=\frac{\delta}{1+\delta}\, M\,,
\eeq
where $\delta\equiv \delta\rho/\rho_0$ is the density contrast.

To study the dynamics of a spherical matter perturbation we need the well known equation
\beq\label{EQ:NonLinearDMDynamics}
\ddot{\delta}-\frac{4}{3}\frac{{\dot{\delta}}^2}{1+\delta}+2H\dot{\delta}=(1+\delta)\,\nabla^2 \Psi\,,
\eeq
which follows from the non-linear continuity and the Euler equation for a pressureless fluid of non-relativistic matter in a top-hat configuration \cite{Schmidt2009f}. 
Eqs. (\ref{EQ:NonLinearEinstein00}), (\ref{EQ:NonLinearEinsteinij}) and (\ref{EQ:NonLinearGalileon}) tell us that, inside a top-hat density perturbation, $\Psi'(r)\propto r$, which means 
that $\nabla^2 \Psi$ will be $r$-independent. Indeed, a top-hat profile, remains a top-hat profile during its whole evolution despite the non-validity of Birkhoff's theorem.


To solve Eq. (\ref{EQ:NonLinearDMDynamics}), we have followed \cite{Schmidt2009f}; here we briefly summarize the main steps. Assuming the total 
mass conservation, $R^3\,\rho_0\,(1+\delta)={\rm const.}$, Eq. (\ref{EQ:NonLinearDMDynamics}) can be rewritten in terms of R
\beq\label{EQ:NonLinearDMDynamicsR}
\frac{\ddot{R}}{R}\,=\,H^2+\dot{H}-\frac{1}{3}\nabla^2\Psi\,.
\eeq
From this equation we can distinguish all the sources that affect the collapse dynamics: $H^2+\dot{H}$ contains the contribution of the background (matter and dark energy), 
while $\nabla^2\Psi$ contains the contribution of matter and scalar field perturbations. Using $N=\ln a$ as a time variable and defining
\beq\label{EQ:ChangeVarRy}
y\equiv\frac{R}{R_i}-\frac{a}{a_i}\,,
\eeq
where $R_i$ and $a_i$ are the initial radius of the perturbation and the initial scale factor, Eq. (\ref{EQ:NonLinearDMDynamicsR}) becomes
\beq\label{EQ:NonLinearDMDynamicsy}
y''+\frac{H'}{H}\, y'-\left(1+\frac{H'}{H}\right)\, y=-\frac{1}{3}\,\left(y+e^{N-N_i}\right)\,\nabla^2\Psi\,,
\eeq
where a prime denotes differentiation w.r.t. $N$. The density contrast is
\beq
\delta=\left(1+\delta_i\right)\cdot{\left(e^{N_i-N}y+1\right)}^{-3}-1\,.
\eeq
Eq. (\ref{EQ:NonLinearDMDynamicsy}) can be solved numerically setting the initial conditions. From Eq. (\ref{EQ:ChangeVarRy}) we know that $y_i=0$ and $y'_i=-\delta'_i/(3(1+\delta_i))$. 
Supposing that the perturbations start growing linearly during matter-dominance, the linearization of Eq. (\ref{EQ:NonLinearDMDynamics}) can be solved analytically. 
The growing mode is $\delta\propto a$, so $\delta'=\delta$, thus the second initial condition becomes $y'_i=-\delta_i/3$. We also set $a_i=10^{-5}$, while the initial density perturbation 
is set to collapse exactly at $a_0=1$.

\subsection{Virialisation}

The Virial Theorem states that a stable system must satisfy the relation
\beq\label{EQ:VirialCondition}
W+2T=0\,,
\eeq
where 
\beq
T\equiv\frac{1}{2}\int d^3 x \rho v^2=\frac{3}{10}M\dot{R}^2
\eeq
is the kinetic energy (the last equality holds true for a top-hat profile), while
\beq
W\equiv-\int d^3 x \rho_m(\vec{x})\vec{x}\cdot\nabla\Psi = -\frac{3 M}{R^3}\sum_i\int_0^R {\rm d}r\cdot r^3 \frac{{\rm d} \Psi_i(r)}{{\rm d} r}
\eeq
is the trace of the potential energy tensor. As in the previous equation the last equality holds true only for a top-hat profile. 
$\Psi_i(r)$ denotes each component that contributes to the total gravitational potential.

Usually energy conservation is used, but, as noted in \cite{Schmidt2010}, for a time-dependent dark energy model, energy is not strictly conserved. 
So, during the collapse phase, the virial radius can be estimated as the radius at which the virial condition (\ref{EQ:VirialCondition}) is satisfied.

Important quantities that can be extrapolated from the dynamics of the collapse are the linearized density contrast $\delta_c$, and the virial overdensity:
\beq
\Delta_{vir}\equiv\frac{\rho_{vir}}{\rho_{collapse}}=\left[1+\delta(R_{vir})\right]\cdot{\left(\frac{a_{collapse}}{a_{vir}}\right)}^3\,.
\eeq

\subsection{Numerical Results}

\subsubsection{Case $\b=0$, $\xds=1$.}

This is the case in which the fifth term of Eq. (\ref{EQ:Action}) gives no contribution. Eqs. (\ref{EQ:IntegralEinstein00}), (\ref{EQ:IntegralEinsteinij}) 
and (\ref{EQ:IntegralGalileon}) become simpler. In particular, the modified Poisson equation reads
\beqra
\nabla^2\Psi&=&3 \Omega_m \Hds^2 a^{-3} x^4 \frac{2x^4-\a}{2x^4+3\a}\,\delta +\nonumber\\
&&-\frac{3 \Hds^2 x^2\left[2x^4(2+\a)+\a(-2+15\a)\right]-36\a (2x^4+3\a)\dot{H}}{\Hds^2 \Mpl {(2x^4+3\a)}^2}\cdot\frac{\varphi'}{r}+\nonumber\\
&&-\frac{12\a x^2 (2x^4-3\a)}{\Hds^2 \Mpl^2 {(2x^4+3\a)}^2}\cdot\frac{\varphi'^2}{r^2}\,,
\eeqra
with $x\equiv H/\Hds$, and $\varphi'$ is a solution of:
\beq
\a_1\cdot\frac{\varphi'^3}{r^3}+\a_2\cdot\frac{\varphi'^2}{r^2}+\left(\a_3+\a_4\delta\right)\cdot\frac{\varphi'}{r}+\a_5\delta=0\,,
\eeq
with:
\beqra
\a_1&=&4 \a x^2\left(4 x^8+24 x^4\a-45\a^2\right)\\
\a_2&=&2 \Mpl\left[\Hds^2 x^2\left(4 x^4(2+3\a)(x^4+6\a)-9\a^2(2-21\a)\right)+\right.\\
\phantom{\a_2=}&&\left.+6\a\left(4x^8-24\a x^4-45\a^2\right)\dot{H}\right]\nonumber\\
\a_3&=&-2 \Hds^2 \Mpl^2 \left[\Hds^2 x^2 \left[2 x^8 (2+\a )+x^4 \left(-4+8 \a +21 \a ^2\right)+\right.\right.\nonumber\\
\phantom{\a_3=}&&\left.\left.+\a  \left(2-21 \a +45 \a ^2\right)\right]-\left[4 x^8 (2+3 \a )+27 \a ^2 (-2+5 \a )+\right.\right.\nonumber\\
\phantom{\a_3=}&&\left.\left.+12 x^4 \a  (-2+9 \a )\right] \dot{H}\right]\\
\a_4&=&-8 e^{-3 n} \Hds^4 \Mpl^2 \Omega_m x^4 \left(2 x^4-3 \a \right) \a\\
\a_5&=&-e^{-3 n} \Hds^4 \Mpl^3 \Omega_m x^2 \left[\Hds^2 x^2 \left(2 x^4 (2+\a )+\a  (-2+15 \a )\right)+\right.\nonumber\\
\phantom{\a_5=}&&\left.-12 \a  \left(2 x^4+3 \a \right) \dot{H}\right]\,.
\eeqra

Of course, among the solutions we want the one that reduces to Eq. (\ref{EQ:LinearSolutionGalileon}) when $r\gg r_V$. 

\begin{figure}[tb]
 \begin{center}
  \includegraphics[width=0.8\textwidth]{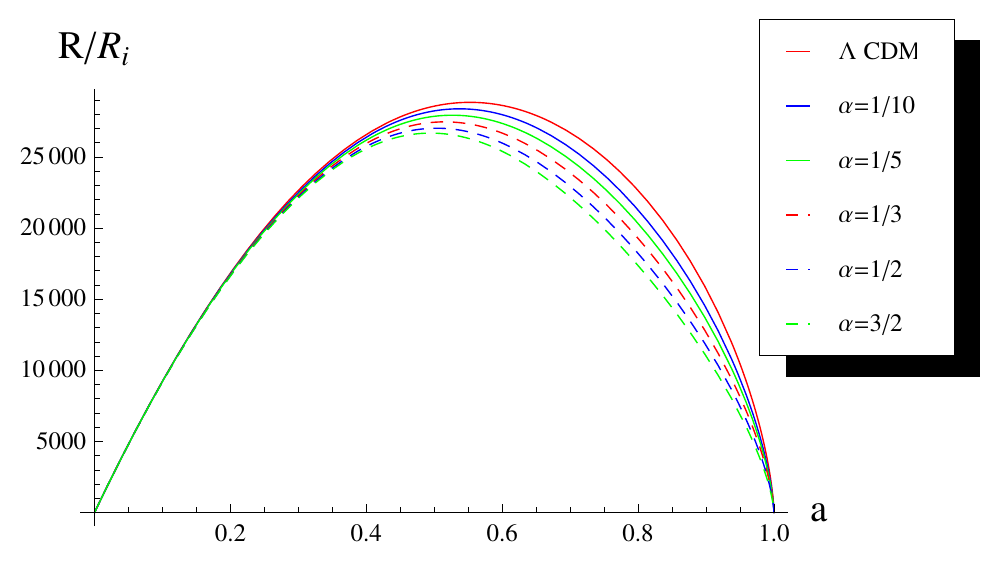}
 \end{center}
 \caption{\textit{In the figure we plot the solution of Eq. (\ref{EQ:NonLinearDMDynamicsy}), in terms of the normalized radius $R/R_i$ of the top-hat perturbation, 
 when $\b=0$ and $\xds=1$. The initial density for each model is shown in Tab. (\ref{TAB:b0}).}}\label{FIG:Collapseb0}
 \end{figure}

Although this is a particular case, it is interesting to show the role of ${\cal L}_4$ in Eq. (\ref{EQ:Action}). In Fig. \ref{FIG:Collapseb0} we have plotted 
the solution of Eq. (\ref{EQ:NonLinearDMDynamicsy}) for various $\a$. It should be noted that modifications w.r.t. the $\lcdm$ model are present 
during the collapse phase. This is, as expected, an effect of the increasing contribution from the scalar field. In Tab. (\ref{TAB:b0}) we show the values 
assumed by the linearized density contrast and the virial overdensity.

\begin{center}
\begin{table}
 \begin{tabular}{|c|c|c|c|c|c|c|c|c|}
  \hline
  Model & $\delta_i$ $(10^{-5})$ & $\delta_c$ & $a_{tur}$ & $R_{tur}/R_i$ & $\Delta_{tur}$ & $a_{vir}$ & $R_{vir}/R_i$ & $\Delta_{vir}$\\
  \hline
  $\lcdm$ & 2.220 & 1.674 & 0.553 & 28840 & 42 & 0.919 & 13910 & 371\\
  $\a=0$ & 2.205 & 1.689 & 0.551 & 28990 & 41 & 0.914 & 14170 & 351\\
  $\a=1/10$ & 2.243 & 1.723 & 0.537 & 28380 & 44 & 0.899 & 14500 & 328\\
  $\a=1/5$ & 2.272 & 1.757 & 0.527 & 27930 & 46 & 0.884 & 14850 & 305\\
  $\a=1/3$ & 2.300 & 1.801 & 0.515 & 27470 & 48 & 0.863 & 15430 & 272\\
  $\a=1/2$ & 2.327 & 1.847 & 0.504 & 27020 & 51 & 0.836 & 16150 & 238\\
  $\a=2/3$ & 2.345 & 1.882 & 0.495 & 26680 & 53 & 0.812 & 16710 & 215\\
  \hline
 \end{tabular}
 \caption{\textit{Here we show numerical results of physical interesting quantities in the case $\b=0$, $\xds=1$ for various $\a$}}\label{TAB:b0}
\end{table}
\end{center}

\subsubsection{Case $\a=0$, $\xds=1$.}

In this paragraph we analyze another particular case, the one which shows the role of ${\cal L}_5$, Eq. (\ref{EQ:Action}), in the dynamics of the collapse. 
Compared to the previous paragraph, when $\b\neq 0$ Eq. (\ref{EQ:NonLinearGalileonEquation}) cannot have an analytic solution. By the parameter conditions, 
Eqs. (\ref{EQ:NoGhostCondition}) and (\ref{EQ:NewParameterConditions}), $-1/3\leq\b\leq 0$, so, to investigate the parameter region in which $\b>0$ we need to set $\a> 0$.

\begin{figure}[tb]
 \begin{center}
  \includegraphics[width=0.8\textwidth]{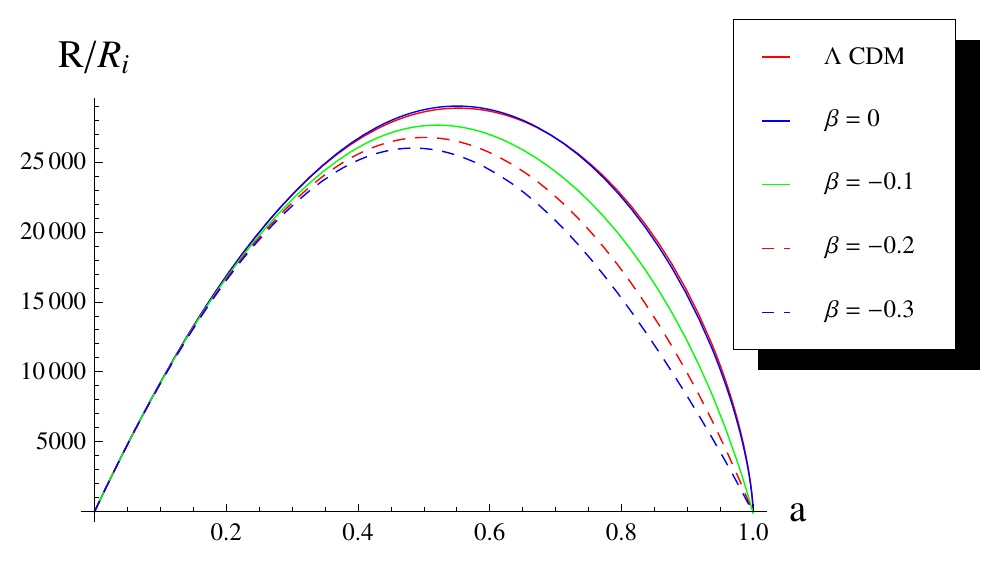}
 \end{center}
 \caption{\textit{In the figure we plot the solution of Eq. (\ref{EQ:NonLinearDMDynamicsy}), in terms of the normalized radius $R/R_i$ of the top-hat perturbation, when $\a=0$ and $\xds=1$. 
 The initial density for each model is shown in Tab. (\ref{TAB:a0}).}}\label{FIG:Collapsea0}
 \end{figure}

The dynamics of the collapse is shown in Fig. \ref{FIG:Collapsea0}, while the linearized density contrast and the virial overdensity for various $\b$ can be found in Table (\ref{TAB:a0}). 
It is important to note that the onset of the fifth term in Eq. (\ref{EQ:Action}) plays a crucial role in the virialisation process. In fact we can see that varying the parameter $\b$ 
there is a substantial modification of $\Delta_{vir}$ with respect to the $\lcdm$ model.

\begin{center}
\begin{table}
 \begin{tabular}{|c|c|c|c|c|c|c|c|c|c|}
  \hline
  Model & $\delta_i$ $(10^{-5})$ & $\delta_c$ & $a_{tur}$ & $R_{tur}/R_i$ & $\Delta_{tur}$ & $a_{vir}$ & $R_{vir}/R_i$ & $\Delta_{vir}$\\
  \hline
  $\lcdm$ & 2.220 & 1.674 & 0.553 & 28840 & 42 & 0.919 & 13910 & 371\\
  $\b=0$ & 2.205 & 1.689 & 0.551 & 28990 & 41 & 0.914 & 14170 & 351\\
  $\b=-0.005$ & 2.219 & 1.700 & 0.547 & 28800 & 42 & 0.907 & 14410 & 334\\
  $\b=-0.01$ & 2.227 & 1.707 & 0.544 & 28680 & 42 & 0.911 & 13810 & 380\\
  $\b=-0.02$ & 2.238 & 1.717 & 0.540 & 28500 & 43 & 0.910 & 13600 & 398\\
  $\b=-0.05$ & 2.263 & 1.742 & 0.531 & 28120 & 45 & 0.895 & 14050 & 361\\
  $\b=-0.07$ & 2.277 & 1.757 & 0.527 & 27910 & 46 & 0.883 & 14470 & 330\\
  $\b=-0.1$ & 2.296 & 1.780 & 0.520 & 27620 & 47 & 0.866 & 15060 & 293\\
  $\b=-0.2$ & 2.356 & 1.857 & 0.501 & 26740 & 52 & 0.813 & 16440 & 225\\
  $\b=-0.3$ & 2.412 & 1.928 & 0.484 & 25980 & 57 & 0.769 & 17150 & 198\\
  \hline
 \end{tabular}
 \caption{\textit{Here we show numerical results of physically interesting quantities, in the case $\a=0$, $\xds=1$ for various $\b$}}\label{TAB:a0}
\end{table}
\end{center}

\subsubsection{Case $\a\neq 0$, $\b\neq 0$, $\xds=1$.}

This is the most general case, despite the assumption $\xds = 1$. Here we can evaluate the sum of the contribution of the terms ${\cal L}_4$ and ${\cal L}_5$, 
Eqs. (\ref{EQ:L4}) and (\ref{EQ:L5}), in the whole parameter region defined by Eq. (\ref{EQ:NewParameterConditions}). It can be noted that as $\a$ and $\b$ 
grow we obtain a larger $\delta_c$, thus it should be easy to remove a large piece of parameter space from the allowed region.

 \begin{figure}[tb]
 \begin{center}
  \includegraphics[width=0.7\textwidth]{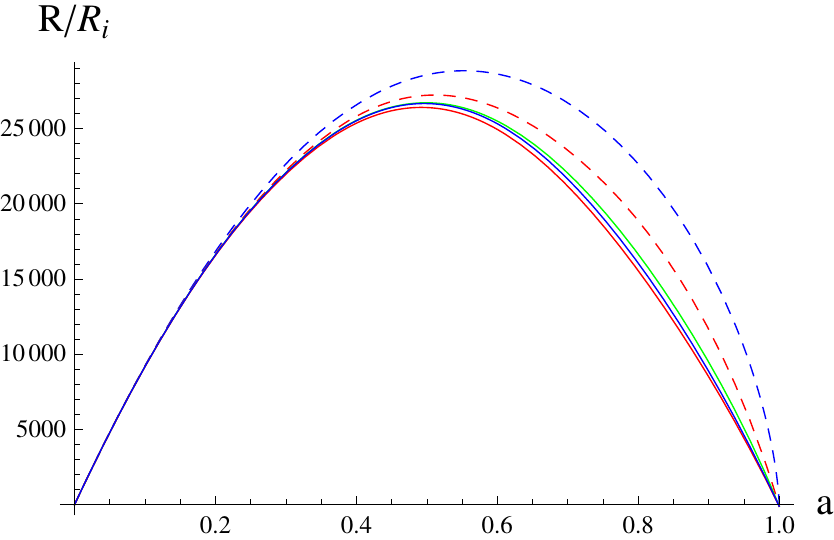}
 \end{center}
 \caption{\textit{In the figure we plot the solution of Eq. (\ref{EQ:NonLinearDMDynamicsy}), in terms of the normalized radius $R/R_i$ of the top-hat perturbation, when 
 $\a\neq 0$, $\b\neq 0$ and $\xds=1$. The initial density for each model is shown in Tab. (\ref{TAB:GeneralCase}). The values for $(\a,\b)$ are: $\lcdm$ blue 
 dashed line; $(-0.45, -0.4)$ red line; $(-0.2, -0.2)$ green line; $(-0.55, -0.4)$ blue thick line; $(0.1, -0.1)$ red dashed line.}}\label{FIG:CollapseGeneralCase}
 \end{figure}

\begin{center}
\begin{table}
 \begin{tabular}{|c|c|c|c|c|c|c|c|c|c|}
  \hline
  \multicolumn{2}{|c|}{Model} & $\delta_i$ $(10^{-5})$ & $\delta_c$ & $a_{tur}$ & $R_{tur}/R_i$ & $\Delta_{tur}$ & $a_{vir}$ & $R_{vir}/R_i$ & $\Delta_{vir}$\\
  \hline
  \multicolumn{2}{|c|}{$\lcdm$} & 2.220 & 1.674 & 0.553 & 28840 & 42 & 0.919 & 13910 & 371\\
  $\a$ & $\b$ & & & & & & & &\\
  $0.1$ & $-0.1$ & 2.323 & 1.815 & 0.511 & 27220 & 50 & 0.862 & 14760 & 311\\
  $-0.2$ & $-0.2$ & 2.356 & 1.831 & 0.499 & 26710 & 52 & 0.791 & 17230 & 195\\
  $-0.45$ & $-0.4$ & 2.383 & 1.875 & 0.492 & 26400 & 54 & 0.763 & 17820 & 177\\
  $-0.55$ & $-0.4$ & 2.362 & 1.851 & 0.496 & 26660 & 53 & 0.773 & 17710 & 180\\
  \hline
 \end{tabular}
 \caption{\textit{Here we show numerical results of physically interesting quantities in the case $\a=0$, $\xds=1$ for various $\b$}}\label{TAB:GeneralCase}
\end{table}
\end{center}


\section{Conclusions}\label{SEC:Conclusions}

In this paper we have first reviewed the background evolution of the Galileon model, following the tracker solution of \cite{DeFelice2010h}. 
We have found two analytic functions that describe the evolution of the components of the universe at late-times. The peculiarity of this tracker solution 
is that it ensures a dS stable point independent of the $c_i$ parameters of Eq. (\ref{EQ:Action}). This assumption simplifies our equations, but it should also be easy to generalize our work to a more general background evolution. Once $c_1$ is set to zero, in Eq. (\ref{EQ:Action}) should remain only kinetic terms for the scalar field, thus the Galileon cannot be considered as a deviation from the $\lcdm$ model. It should work as a substitute of the cosmological constant, mimicking the effects of $\Lambda$ on cosmological scales.\\
Then we have shown that, in the linear approximation the scalr perturbations of a FRLW universe lead to a time-dependent gravitational constant $G_\phi(t)$, that modifies 
the gravitational potential generated by a distant or, equivalently, small source. The results we give do not represent a realistic model, i.e. they are not required to satisfy the
observational bounds, rather they are chosen on order to display what one can generally expect from this theory.\\
The Galileon model can be successful because it possesses a Vainshtein mechanism, by which we can consider two distinct regions; the first one at large scales, where the linear approximation applies
and the Galileon drives the cosmic acceleration, the second one where non-linearities are dominant. 
We have also shown how to recover a Vainshtein radius in agreement with the one of DGP and other simpler models.\\
Even though the study of the perturbations in a highly non-linear regime can notbe completely analytic, we found some constraints, whose fulfillment allows Eq. (\ref{EQ:ApproxNonLinearGalileon}) 
to have at least a couple of real solutions.\\
The last part of this paper was devoted to the study of the collapse of a spherical top-hat matter perturbation. We have shown that the new terms ${\cal L}_4$ and ${\cal L}_5$ affect in a non-negligible 
way the dynamics of the collapse and the value of $\delta_c$ and $\Delta_{vir}$. To study the virialisation process we paid attention to the energy non-conservation problem, calculating 
point by point the virial condition Eq. (\ref{EQ:VirialCondition}). 

\section*{Acknowledgments}
We thank Daniele Bertacca, Bin Hu, Massimo Pietroni, Jos\'e M. Mart\'{i}n-Garc\'{i}a and the community of the \textsc{Mathematica} package \textit{xAct} for useful discussions.

\appendix

\section{Components of the field equations (\ref{EQ:Einstein}-\ref{EQ:Galileon})} \label{Appendix:EOMComponents}

The terms of the stress-energy tensor of the scalar field read
\beqra \label{EQ:EnergyStressComponents}
T^{(1)}_{\phantom{(1)}\mu\nu}&=&\frac{1}{2} M^3 g_{\mu\nu}\,\phi \\
T^{(2)}_{\phantom{(2)}\mu\nu}&=&-\phi_{;\mu}\, \phi_{;\nu}+\frac{1}{2} g_{\mu\nu}\, (\nabla \phi)^2  \\
T^{(3)}_{\phantom{(3)}\mu\nu}&=&-\frac{1}{M^3}\left[\phi_{;\mu}\,\phi_{;\nu}\, \square\phi-\phi_{;\{\mu}\,\phi_{;\nu\}\a}\,\phi^{;\a}+g_{\mu\nu}\,\phi^{;\a}\,\phi_{;\a\b}\,\phi^{;\b}\right] \\
T^{(4)}_{\phantom{(4)}\mu\nu}&=&-\frac{2}{M^6}\left[-\frac{1}{2} R\, \phi_{;\mu}\, \phi_{;\nu}\, (\nabla \phi)^2 +2 \phi_{;\mu\a}\, \phi^{;\a}\, \phi_{;\nu\b}\, \phi^{;\b} +\right.\nonumber\\
& &\left. -2 \phi_{;\mu\nu}\, \phi^{;\a}\, \phi_{;\a\b}\, \phi^{;\b}+2 \phi_{;\{\mu}\, \phi_{;\nu\}\a}\, \phi_{;\b}\, \phi^{;\b\a} -\phi_{;\mu}\, \phi_{;\nu}\, \phi_{;\a\b}\, \phi^{;\a\b} +\right.\nonumber\\
& &\left.+R_{\a\{\mu}\, \phi_{;\nu\}}\, \phi^{;\a}\, (\nabla \phi)^2-\frac{1}{4} G_{\mu\nu} (\nabla \phi)^4 + \phi_{;\mu\a}\, \phi_{;\nu}^{\phantom{;\nu}\a}\, (\nabla \phi)^2 +\right.\nonumber\\
& &\left.-g_{\mu\nu}\, R_{\a\b}\, \phi^{;\a}\, \phi^{;\b}\, (\nabla \phi)^2+R_{\mu\a\nu\b}\, \phi^{;\a}\, \phi^{;\b}\, (\nabla \phi)^2 +\right.\nonumber\\
& &\left.-2 g_{\mu\nu}\, \phi_{;\a}\, \phi^{;\b}\, \phi_{;\b\g}\, \phi^{;\a\g}-\phi_{;\mu\nu}\, \square\phi\, (\nabla \phi)^2 -2 \phi_{;\{\mu}\, \phi_{;\nu\}\a}\, \phi^{;\a}\, \square\phi +\right.\nonumber\\
& &\left.+\phi_{;\mu}\, \phi_{;\nu}\, (\square\phi)^2-\frac{1}{2} g_{\mu\nu}\, \phi_{;\a\b}\, \phi^{;\a\b}\, (\nabla \phi)^2+2 g_{\mu\nu}\, \phi^{;\a}\, \phi_{;\a\b}\, \phi^{;\b}\, \square\phi+ \right.\nonumber\\
& &\left.+\frac{1}{2} g_{\mu\nu}\, (\square\phi)^2\, (\nabla \phi)^2 \right]\\
T^{(5)}_{\phantom{(5)}\mu\nu}&=&-\frac{2}{M^9}\left[3 \phi_{;\mu\g}\, \phi_{;\nu}^{\phantom{;\nu}\g}\, \phi^{;\a}\, \phi_{;\a\b}\, \phi^{;\b} -3 \phi_{;\mu\a}\, \phi_{;\nu\g}\, \phi^{;\g}\, \phi_{;\b}\, \phi^{;\b\a}+\right.\nonumber\\
& &\left.+ \frac{3}{2} \phi_{;\mu\nu}\, R_{\a\b}\, \phi^{;\a}\, \phi^{;\b}\, (\nabla \phi)^2 -\frac{3}{2} R_{\a\{\mu}\, \phi_{;\nu\}\b}\, \phi^{;\a}\, \phi^{;\b}\, (\nabla \phi)^2 +\right.\nonumber\\
& &\left.+\frac{3}{4} R\, \phi_{;\{\mu}\, \phi_{;\nu\}\a}\, \phi^{;\a}\, (\nabla \phi)^2 +\frac{3}{2} G_{\mu\nu}\, \phi^{;\a}\, \phi_{;\a\b}\, \phi^{;\b}\, (\nabla \phi)^2 +\right.\nonumber\\
& &\left.-\frac{3}{2} R_{\b}^{\phantom{\b}\a}\, \phi_{;\{\mu}\, \phi_{;\nu\}\a}\, \phi^{;\b}\, (\nabla \phi)^2 -3 \phi_{;\mu\a}\, \phi_{;\nu\b}\, \phi^{;\a}\, \phi_{;\g}\, \phi^{;\b\g} +\right.\nonumber\\
& &\left.-3 \phi_{;\mu\a}\, \phi^{;\a\b}\, \phi_{;\nu\b}\, (\nabla \phi)^2 +3 \phi_{;\mu\nu}\, \phi_{;\a}\, \phi^{;\a\b}\, \phi_{;\g\b}\, \phi^{;\g} +\right.\nonumber\\
& &\left.-3 \phi_{;\{\mu}\, \phi_{;\nu\}\a}\, \phi^{;\a\b}\, \phi_{;\g\b}\, \phi^{;\g}+\frac{3}{2} R_{\a\b}\, \phi_{;\mu}\, \phi_{;\nu}\, \phi^{;\a\b}\, (\nabla \phi)^2 +\right.\nonumber\\
& &\left.+\phi_{;\mu}\, \phi_{;\nu}\, \phi_{;\a}^{\phantom{;\a}\b}\, \phi_{;\b\g}\, \phi^{;\a\g}+\frac{3}{4} R\, \phi_{;\mu}\, \phi_{;\nu}\, \square\phi\, (\nabla \phi)^2 +\right.\nonumber\\
& &\left.+\frac{3}{2} \phi_{;\mu\nu}\, \phi_{;\a\b}\, \phi^{;\a\b}\, (\nabla \phi)^2 +\frac{3}{2} \phi_{;\{\mu}\, \phi_{;\nu\}\a}\, \phi^{;\a}\, \phi_{;\b\g}\, \phi^{;\b\g} +\right.\nonumber\\
& &\left. -\frac{3}{2} \phi_{;\mu}\, \phi_{;\nu}\, \phi_{;\a\b}\, \phi^{;\a\b}\, \square\phi+\frac{3}{2} R_{\a \g \b\{\mu}\, \phi_{;\nu\}}^{\phantom{;\nu\}}\a}\, \phi^{;\b}\, \phi^{;\g}\, (\nabla \phi)^2 +\right.\nonumber\\
& &\left.-\frac{3}{2} R_{\a\{\mu}\, \phi_{;\nu\}}\, \phi_{;\b}\, \phi^{;\b\a}\, (\nabla \phi)^2-\frac{3}{2} R_{\a\g\b\{\mu}\, \phi_{;\nu\}}\, \phi^{;\g}\, \phi^{;\a\b}\, (\nabla \phi)^2 +\right.\nonumber\\
& &\left. -\frac{3}{2} R\, \phi_{;\mu}\, \phi_{;\nu}\, \square\phi\, (\nabla \phi)^2+\frac{3}{2} R_{\a\{\mu}\, \phi_{;\nu\}}\, \phi^{;\a}\, \square\phi\, (\nabla \phi)^2 +\right.\nonumber\\
& &\left. +3 g_{\mu\nu}\, \phi_{;\a}\, \phi_{;\b}\, \phi^{;\a\g}\, \phi_{;\g\tau}\, \phi^{;\b\tau}+3 g_{\mu\nu}\, R_{\g\b}\, \phi_{;\a}\, \phi^{;\g}\, \phi^{;\a\b}\, (\nabla \phi)^2 +\right.\nonumber\\
& &\left. -\frac{3}{2} R_{\mu\nu\a\b}\, \phi_{;\g}\, \phi^{;\a}\, \phi^{;\b\g}\, (\nabla \phi)^2-3 R_{\mu\b\nu\g}\, \phi_{;\a}\, \phi^{;\g}\, \phi^{;\a\b}\, (\nabla \phi)^2 +\right.\nonumber\\
& &\left. +\frac{3}{2} g_{\mu\nu}\, R_{\a\g\b\tau}\, \phi^{;\a}\, \phi^{;\b}\, \phi^{;\g\tau}\, (\nabla \phi)^2+g_{\mu\nu}\, \phi^{;\a}_{\phantom{;\a}\b}\, \phi_{;\a\g}\, \phi^{;\b\g}\, (\nabla \phi)^2 +\right.\nonumber\\
& &\left. +3 \phi_{;\mu\a}\, \phi^{;\a}\, \phi_{;\nu\b}\, \phi^{;\b}\, \square\phi -3 \phi_{;\mu\nu}\, \phi^{;\a}\, \phi_{;\a\b}\, \phi^{;\b}\, \square\phi+\right.\nonumber\\
& &\left. +3 \phi_{;\{\mu}\, \phi_{;\nu\}\a}\, \phi^{;\b\a}\, \phi_{;\b}\, \square\phi+3 \phi_{;\mu\a}\, \phi_{;\nu}^{\phantom{;\nu}\a}\, \square\phi\, (\nabla \phi)^2 +\right.\nonumber\\
& &\left.-\frac{3}{2} g_{\mu\nu}\, R_{\a\b}\, \phi^{;\a}\, \phi^{;\b}\, \square\phi\, (\nabla \phi)^2 +\frac{3}{2} R_{\mu\a\nu\b}\, \phi^{;\a}\, \phi^{;\b}\, \square\phi\, (\nabla \phi)^2 +\right.\nonumber\\
& &\left.-\frac{3}{2} \phi_{;\mu\nu}\, (\square\phi)^2\, (\nabla \phi)^2 -\frac{3}{2} \phi_{;\{\mu}\, \phi_{;\nu\}\a}\, \phi^{;\a}\, (\square\phi)^2+\frac{1}{2} \phi_{;\mu}\, \phi_{;\nu}\, (\square\phi)^3 +\right.\nonumber\\
& &\left.-\frac{3}{2} g_{\mu\nu}\, \phi^{;\a}\, \phi_{;\a\b}\, \phi^{;\b}\, \phi_{;\g\tau}\, \phi^{;\g\tau}-\frac{3}{2} g_{\mu\nu}\, \phi_{;\a\b}\, \phi^{;\a\b}\, \square\phi\, (\nabla \phi)^2+\right.\nonumber\\
& &\left.-3 g_{\mu\nu}\, \phi_{;\a}\, \phi^{;\g}\, \phi_{;\g\b}\, \phi^{;\a\b}\, \square\phi +\frac{3}{2} g_{\mu\nu}\, \phi^{;\a}\, \phi_{;\a\b}\, \phi^{;\b}\, (\square\phi)^2+\right.\nonumber\\
& &\left.+\frac{1}{2} g_{\mu\nu} (\square\phi)^3 (\nabla \phi)^2 \right].
\eeqra

The terms appearing in the equation of motion for the scalar field read
\beqra
\xi^{(1)}&=&\frac{M^3}{2} \\
\xi^{(2)}&=&-\square\phi \\
\xi^{(3)}&=&\frac{1}{M^3} \left[-(\square\phi)^2 +R_{\mu\nu}\, \phi^{;\mu}\, \phi^{;\nu} +\phi_{;\mu\nu}\, \phi^{;\mu\nu} \right] \\
\xi^{(4)}&=&\frac{1}{M^6}\left[2 R\, \phi^{;\mu}\, \phi_{;\mu\nu}\, \phi^{;\nu} -8 R_{\nu\a}\, \phi^{;\mu}\, \phi^{;\nu}\, \phi_{;\mu}^{\phantom{;\mu}\a} -2 R_{\mu\nu}\, \phi^{;\mu\nu}\, (\nabla \phi)^2 +\right.\nonumber\\
& &\left.-4 R_{\mu\a\nu\b}\, \phi^{;\mu}\, \phi^{;\nu}\, \phi^{;\a\b} -4 \phi_{;\mu}^{\phantom{;\mu}\nu}\, \phi_{;\nu}^{\phantom{;\nu}\a}\, \phi^{;\mu}_{\phantom{;\mu}\a} +R\, (\nabla \phi)^2\, \square\phi +\right.\nonumber\\
& &\left. +4 R_{\mu\nu}\, \phi^{;\mu}\, \phi^{;\nu}\, \square\phi +6 \phi_{;\mu\nu}\, \phi^{;\mu\nu}\, \square\phi -2 (\square\phi)^3 \right] \\
\xi^{(5)}&=&\frac{1}{M^9}\left[\frac{3}{2}  R\, (\nabla \phi)^2\, (\square\phi)^2 +3  R\, \phi^{;\mu}\, \phi_{;\mu\nu}\, \phi^{;\nu}\, \square\phi +\right.\nonumber\\
& &\left.+3 R_{\mu}^{\phantom{\mu}\a}\, R_{\nu\a}\, \phi^{;\mu}\, \phi^{;\nu}\, (\nabla \phi)^2-\frac{3}{2} R\, R_{\mu\nu}\, \phi^{;\mu}\, \phi^{;\nu}\, (\nabla \phi)^2 +\right.\nonumber\\
& &\left. +3  R^{\mu\nu}\, R_{\a\mu\b\nu}\, \phi^{;\a}\, \phi^{;\b}\, (\nabla \phi)^2-\frac{3}{2} R_{\mu}^{\phantom{\mu}\a\b\g}\, R_{\nu\a\b\g}\, \phi^{;\mu}\, \phi^{;\nu}\, (\nabla \phi)^2 +\right.\nonumber\\
& &\left.-3  R\, \phi^{;\mu}\, \phi^{;\nu}\, \phi_{;\mu\a}\, \phi_{;\mu}^{\phantom{;\mu}\a}-\frac{3}{2} R\, \phi_{;\mu\nu}\, \phi^{;\mu\nu}\, (\nabla \phi)^2- (\square\phi)^4 +\right.\nonumber\\
& &\left. +3 R_{\mu\nu}\, \phi^{;\mu}\, \phi^{;\nu}\, (\square\phi)^2-12  R_{\mu\a}\, \phi^{;\mu}\, \phi^{;\nu}\, \phi_{;\nu}^{\phantom{;\nu}\a}\,  \square\phi +\right.\nonumber\\
& &\left.+6 R_{\a\b}\, \phi^{;\mu}\, \phi^{;\nu}\, \phi_{;\mu}^{\phantom{;\mu}\a}\, \phi_{;\nu}^{\phantom{;\nu}\b}+6 R_{\mu\nu}\, \phi^{;\mu\a}\, \phi_{;\a}^{\phantom{;\a}\nu}\, (\nabla \phi)^2 +\right.\nonumber\\
& &\left.  +6 \phi^{;\mu\nu}\, \phi_{;\mu\a}\, \phi^{;\a\b}\, \phi_{;\nu\b}-8 \phi^{;\mu\nu}\, \phi_{;\nu\a}\, \phi_{;\mu}^{\phantom{;\mu}\a}\, \square\phi +\right.\nonumber\\
& &\left.+12 R_{\nu\b}\, \phi^{;\mu} \phi^{;\nu}\, \phi_{;\mu\a}\, \phi^{;\a\b} -6 R_{\mu\nu}\, \phi^{;\mu\nu}\, (\nabla \phi)^2\, \square\phi +\right.\nonumber\\
& &\left. -6 R_{\mu\nu}\, \phi^{;\mu\nu}\, \phi^{;\a}\, \phi_{;\a\b}\, \phi^{;\b}+6 \phi_{;\mu\nu}\, \phi^{;\mu\nu}\, (\square\phi)^2 -3  R_{\a\b}\, \phi^{;\a}\, \phi^{;\b}\, \phi_{;\mu\nu}\, \phi^{;\mu\nu}+\right.\nonumber\\
& &\left.-3 (\phi_{;\mu\nu}\, \phi^{;\mu\nu})^2+6  R_{\mu\a\nu\b}\, \phi^{;\mu}\, \phi^{;\nu}\, \phi^{;\g\a}\, \phi_{;\g}^{\phantom{;\g}\b} -6  R_{\mu\a\nu\b}\, \phi^{;\mu}\, \phi^{;\nu}\, \phi^{;\a\b}\, \square\phi +\right.\nonumber\\
& &\left.+12  R_{\mu\a\nu\b}\, \phi^{;\g}\, \phi^{;\mu}\, \phi_{;\g}^{\phantom{;\g}\nu}\, \phi^{;\a\b}+3 R_{\mu\a\nu\b}\, \phi^{;\mu\nu}\, \phi^{;\a\b}\, (\nabla \phi)^2 \right]
\eeqra

\section{Background functions} \label{Appendix:BackgroundFunctions}

Background functions involved in the linear perturbation theory:
\beqra
\g_1(t)&\equiv& 3 \left(\a-2 \xds\b\right) \frac{\dphi^2}{\Hds^4 \Mpl^2} \\
\g_2 (t)&\equiv& \left(2+9 \a -9 \b-12\xds\a+15\xds^2\b\right)\frac{\dphi^2}{\Hds^2 \Mpl}\\
\g_3 (t)&\equiv& -\frac{\dphi^4}{3 \Hds^4 \Mpl^2}\left(\a+6\b\frac{\ddphi}{\Hds^2 \Mpl}\right)\\
\g_4 (t)&\equiv& -\frac{2 \dphi^2}{3 \Hds^2 \Mpl} \left(2 \a-3 \xds\b\right) \left(\xds+\frac{3 \ddphi}{\Hds^2 \Mpl}\right) \\
\g_5 (t)&\equiv& -6-9\a+12\b-26\xds^2\a+4\xds \left(2+9\a-9\b\right)+24\xds^3\b+\nonumber\\
&&+2\left[2+9\a-9\b-6\xds\left(\a-\xds\b\right)\right]\frac{\ddphi}{\Hds^2 \Mpl}
\eeqra
Background functions involved in the non-linear dynamics:
\beqra
\e_1(t)&\equiv&\frac{2 \b}{\Hds^6 \Mpl^3}\dphi^2\\
\e_2(t)&\equiv& \frac{\dphi^2}{3 \Hds^4 \Mpl^2}\left(\a-6\b\frac{\ddphi}{\Hds^2 \Mpl}\right)\\
\e_3(t)&\equiv&\frac{1}{\Hds^2 \Mpl}\left[2+9 \a -9 \b-6 \left(\a-\xds\b\right)\left(\xds+\frac{\ddphi}{\Hds^2 \Mpl}\right)\right]\\
\e_4(t)&\equiv&\frac{2}{\Hds^4 \Mpl^2}\left(\a-2\b\frac{\ddphi}{\Hds^2 \Mpl}\right)
\eeqra

\section*{References}
 \bibliographystyle{JHEP}
 \bibliography{biblio}
\end{document}